\documentclass[twocolumn,secnumarabic,amssymb, nobibnotes, aps, prd]{revtex4-1}

\usepackage{graphicx}
\usepackage{wrapft}
\usepackage{amsmath}
\usepackage{bm}
\usepackage{mathrsfs}
\usepackage{color}

\begin{document}

\title{Magnetic Properties of Quantized Vortices in Neutron 
$^3P_2$ Superfluids\\ in Neutron Stars}

\author{Kota Masuda$^{1,2}$ and Muneto Nitta$^{3}$}
\email{masuda(at)nt.phys.s.u-tokyo.ac.jp, 
nitta(at)phys-h.keio.ac.jp}
\affiliation{$^{1}$Department of Physics, The University of Tokyo, Tokyo 113-0033, Japan\\
$^{2}$Theoretical Research Division, Nishina Center, RIKEN, Wako 351-0198, Japan\\
$^{3}$Department of Physics at Hiyoshi, and Research and Education Center for Natural Sciences,\\ Keio University,
Hiyoshi 4-1-1, Yokohama, Kanagawa 223-8521, Japan}
\date{\today}

\begin{abstract}
We discuss quantized vortices in neutron $^3P_2$ superfluids,
which are believed to realize in high density neutron matter such as neutron stars.
By using the Ginzburg-Landau free energy for $^3P_2$ superfluids, 
we determine the ground state 
in the absence and presence of the external magnetic field, 
and  
numerically construct $^3P_2$ quantized vortices 
in the absence and presence of the external magnetic field 
along the vortex axis (poloidal) or angular direction (toroidal). 
We find  in certain situations  
the spontaneous magnetization of the vortex core, 
whose typical magnitude is about $10^{7-8}$ Gauss,
but the net magnetic field 
in a neutron star is negligible because of 
the ratio of the 
vortex core size $\sim 10$fm 
and the intervortex distance $\sim 10^{-6}$m 
in a vortex lattice. 

\end{abstract}

\maketitle

\section{Introduction}

Neutron stars provide unique laboratories in universe 
not only for astrophysics but also for nuclear physics and condensed matter physics.
Neutron stars have some observables such as the mass 
($M$), radius ($R$), 
surface temperature ($T_s$) and magnetic fields on the surface ($B_s$).
The observed masses of neutron stars 
give a stringent constraint on the stiffness of the equation of state.
Other observables such as $T_s$ and $B_s$ give us 
rich information about the states of high density nuclear matter.

It is generally believed that a neutron superfluid state is realized inside neutron stars, 
which is a high density fermionic system.
At low densities less than the normal nuclear matter density, the dominant effective pair interaction is the $^1S_0$ attractive one, 
and the possibility of $^1S_0$ superfluidity was pointed out by Migdal in 1959 
\cite{Migdal:1959}. 
From the observational point of view,
pulsar glitches, which are 
the sudden speed-up events of neutron stars \cite{Reichley1971}, 
might show the existence of the superfluidity inside neutron stars, 
although the mechanism of pulsar glitches is still controversial.
The origin of pulsar glitches was proposed to be 
the starquake from the core or the crust of neutron stars \cite{glitch-1,glitch-2,glitch-3}.
It was also proposed that pulsar glitches can be explained by 
the unpinning dynamics of a large number of neutron vortices
pinned on the nuclei \cite{Anderson:1975zze}.
In spite of several different proposals,
there is a common point for the existence of superfluids among several models; 
The observed long relaxation time $\tau$ ($\sim$ weeks for Crab and $\sim$ years for Vela)
can be explained by assuming that neutron stars have the two components, normal neutrons and superfluid neutrons \cite{glitch-1,glitch-2,glitch-3}. 
Moreover, recent observation of the cooling process of a neutron star 
may indicate the existence of superfluid components in the neutron star
\cite{Heinke:2010cr,Page:2010aw}. 
Therefore,
it is very important to study the detailed properties of neutron superfluidity 
to understand the dynamics and evolution of neutron stars. 
Once it is established,  a large number of 
quantized vortices are inevitably created along the rotation axis 
due to the rapid rotation of neutron stars, 
and consequently understanding the dynamics of superfluid vortices 
should be crucial. 

Due to the repulsive core in the $^1S_0$ partial wave, 
the effects of pairing in the $^3P_2$ attractive interaction becomes comparable to that of the $^1S_0$ case at about the normal nuclear matter density.  
Therefore, a transition from an isotropic $^1S_0$ superfluid to an anisotropic $^3P_2$ superfluid has been predicted to occur at this density 
\cite{tamagaki:1970,PhysRevLett.24.775,takatsuka:1971,takatsuka:1972,
fujita:1972,Richardson:1972xn}.
The Ginzburg-Landau (GL) free for the $^3P_2$ superfluid, 
which is valid near the critical temperature, 
was derived in Refs.~\cite{fujita:1972,Richardson:1972xn} 
in the weak coupling limit. 
The ground state was determined in the GL theory
 to be in the nematic phase \cite{Sauls:1978lna}
according to the classification by Mermin \cite{Mermin:1974zz} 
for the GL free energy with total angular momentum two. 
The strong coupling effect was taken into account in Ref.~\cite{Vulovic:1984kc}.
Although no definite observational signal of the existence of 
$^3P_2$ superfluids was obtained yet, 
the existence of quantized vortices is inevitable if it is realized. 
Vortex structures in $^3P_2$ superfluids were discussed 
in the GL equation for the $^3P_2$ superfluid 
\cite{Richardson:1972xn,Muzikar:1980as,Sauls:1982ie}. 
In particular, the spontaneous magnetization of a vortex 
was pointed out in Ref.~\cite{Sauls:1982ie}. 
Recent study of $^3P_2$ superfluids includes for instance 
low energy excitations, 
their low energy theory, 
neutrino emission 
\cite{Bedaque:2003wj,Bedaque:2012bs,Bedaque:2014zta,
Leinson:2009nu,Leinson:2010yf,Leinson:2010pk,Leinson:2010ru,Leinson:2011jr,Leinson:2012pn,Leinson:2013si},
their effects on cooling process \cite{Leinson:2014cja,Leinson:2015pca}
and the entrainment \cite{Shahabasyan:2011zz}.

In this paper, 
we determine the ground state in the presence of magnetic fields, and 
work out quantized vortex structures in the $^3P_2$ superfluids 
in the presence and absence of magnetic fields  
in the framework of the GL theory.
Due to the tensorial nature of the order parameter 
of the $^3P_2$ superfluids, 
physics depends on the basis in which  
the tensor order parameter take a form.
The vortex was studied before in the absence of the external magnetic field 
and the sixth order term in the GL free energy, 
in which case 
the coordinate basis of the tensor order parameter is cylindrical for 
the vortex solution with the least energy 
\cite{Richardson:1972xn,Muzikar:1980as,Sauls:1982ie}.
We obtain the full numerical solution in this case.
We further take into account the effect of the sixth order term 
in the absence and presence of the magnetic field. 
We find that 
the Cartesian ($xyz$) basis for the tensor order parameter 
give the least energy configurations 
in the presence of the sixth order term and/or 
the magnetic field along the vortex axis while 
the cylindrical basis are preferred only in the absence of 
the external magnetic fields or 
in the presence of the magnetic field along the angular 
direction encircling the vortex. 
We construct the vortex profiles in all these cases.  
We further calculate the magnetization of the vortex core 
induced by the neutron anomalous magnetic moment
and find that it is present only when 
off-diagonal elements in the tensor order parameter 
appear around the vortex core;
the case that the off-diagonal elements have the same winding number
with the diagonal elements 
when the cylindrical basis is preferred 
for the tensor order parameter, 
and the case that 
the off-diagonal elements have a winding number 
differed from that of the diagonal elements by two  
when the Cartesian basis is preferred 
for the tensor order parameter.
Among these,  
a net magnetization is present for the former case, 
that is, the case that 
the sixth order term is negligible 
in the absence the magnetic field
and the case in the presence of the magnetic fields 
in the angular direction. 
For the Cartesian basis, we find a magnetization 
upward and downward along the vortex axis 
locally existing in the angular coordinate,
with zero net magnetization.  
In these cases, 
the typical magnitude of the magnetic field inside the vortex core 
is about $10^{7-8}$ Gauss,
and the average value is much less when averaged in the vortex lattice.

This paper is organized as follows. 
In Sec.~\ref{sec:GL}
we introduce the GL equation for $^3P_2$ superfluid states and 
determine the ground states in the absence and presence of the magnetic fields. 
In the weak coupling limit, degenerate ground states can be parameterized by one parameter.
We also point out some analogy to that of $^3$He superfluids 
and spin-2 Bose-Einstein condensates (BEC). 
In Sec.~\ref{sec:vortex} we construct vortex solutions numerically 
by using the $^3P_2$ GL equation 
in various cases in the presence and absence of magnetic fields
along the vortex axis or angular direction.
In Sec.~\ref{sec:magnetization}, 
we calculate a spontaneous magnetization caused by the $^3P_2$ vortices.
Sec.~\ref{sec:summary} is devoted to a summary and discussion.
In Appendix \ref{sec:ground_state} we give the detailed calculation 
to determine the ground states with taking into account the sixth order term.
In Appendix \ref{sec:eom} we give full equations of motion 
of $^3P_2$ superfluids.

\section{Ginzburg Landau Free Energy for $^3P_2$ Superfluids}\label{sec:GL}

In this section, we first give the GL free energy 
and determine the ground states in various cases
 in the absence and presence of the magnetic fields.

\subsection{Ginzburg-Landau free energy}

The GL free energy for the $^3P_2$ superfluidity in the weak coupling limit 
was derived in 
Refs.~\cite{fujita:1972,Richardson:1972xn, sauls-thesis} assuming 
the contact interaction.
Here let us follow their derivation. 
To this end, we consider properties of dense neutron matter by the following Hamiltonian $H$,
which includes a zero range $^3P_2$ force
\begin{eqnarray}
H=\int d^3 \rho\ \psi^{\dagger}\left(-\frac{\bm{\nabla}^2}{2M}-\mu\right)\psi
-\frac{1}{2}gT^{\dagger}_{\alpha \beta}(\bm{\rho})T_{\alpha \beta}(\bm{\rho})
\label{hamiltonian}
\end{eqnarray}  
where $\bm{\rho}$ denote space coordinates, $\psi$ is a neutron field, $\mu$ is a baryon chemical potential, 
$M$ is the mass of neutrons, and $g(>0)$ is the coupling constant. 
Here, $\alpha,\beta$ are the space indices, and 
the tensor $T_{\alpha\beta}$ is given by 
\begin{eqnarray}
  T^{\dagger}_{\alpha \beta} (\bm{\rho}) 
 = \psi^{\dagger}_{\sigma}(\bm{\rho})
    (t^{\ast}_{\alpha \beta})_{\sigma \sigma'}(\bm{\nabla})
    \psi^{\dagger}_{\sigma'}(\bm{\rho}) 
\end{eqnarray} 
with a differential operator $t$ defined by
\begin{eqnarray}
(t_{\alpha \beta})_{\sigma\sigma'}(\bm{\nabla})
&=&\frac{1}{2}((S_{\alpha})_{\sigma\sigma'}\nabla_{\beta}
+\nabla_{\alpha}(S_{\beta})_{\sigma\sigma'}) \nonumber \\
&&-\frac{1}{3}\delta_{\alpha \beta}(\bm{S})_{\sigma\sigma'}\cdot \bm{\nabla}
\end{eqnarray}
and $\bm{S}$ defined by $(S_{\alpha})_{\sigma\sigma'} = 
i(\sigma_y \sigma_{\alpha})_{\sigma\sigma'}$ $(\alpha=x,y,z)$.

The order parameter for $^3P_2$ superfluidity is  $3 \times 3$ traceless symmetric tensor $A_{\mu i}$, which is defined by
\begin{eqnarray}
\Delta=\sum_{\mu i} i\sigma_{\mu}\sigma_y A_{\mu i} k_i
\end{eqnarray}
where $\Delta$ is the gap parameter.
The Latin letter $\mu$ stands for the spin index as before 
while the Roman index $i$ stands for the spatial coordinates.
The symmetry acts on the tensor $A_{\mu i}$ as
\begin{eqnarray}
 A \to e^{i \theta} g A g^T, \quad e^{i \theta} \in U(1),
\quad g \in SO(3)
\end{eqnarray}
in the matrix notation. 
The free energy density $F$ as a function of tensor $A_{\mu i}$ can be written as
\begin{eqnarray}
F=\int d^3 \rho \ (f_{\rm grad} + f_{2+4}+f_6+f_H) \label{freeenergydensity}
\end{eqnarray}
where $f_{\rm grad}$ is the gradient term, 
$f_{2+4}$ and $f_6$  \cite{sauls-thesis} are the free energy densities up to fourth order and of the sixth order, 
respectively, 
 and $f_H$ is the magnetic term, given by 
\begin{eqnarray}
 f_{\rm grad} &=& K_1 \partial_iA_{\mu j}\partial_iA^{\dagger}_{\mu j} 
 + K_2(\partial_iA_{\mu i}\partial_jA^{\dagger}_{\mu j}+\partial_iA_{\mu j}\partial_jA^{\dagger}_{\mu i})  \nonumber \\  \\
f_{2+4}&=&\alpha {\rm Tr}AA^{\dagger}
 +\beta[({\rm Tr}AA^{\dagger})^2-{\rm Tr}A^2A^{\dagger 2}],\\
f_6 &=&
\gamma [-3({\rm Tr}AA^{\dagger})|{\rm Tr}AA|^2
+4({\rm Tr}AA^{\dagger})^3 \nonumber \\
&&+12({\rm Tr}AA^{\dagger}){\rm Tr}(AA^{\dagger})^2
+6({\rm Tr}AA^{\dagger}){\rm Tr}(A^2A^{\dagger 2}) \nonumber \\ 
&&
+8{\rm Tr}(AA^{\dagger})^3
+12{\rm Tr}[(AA^{\dagger})^2A^{\dagger}A] \nonumber \\
&&-12{\rm Tr}[AA^{\dagger}A^{\dagger}A^{\dagger}AA]
-12{\rm Tr}AA({\rm Tr}AA^{\dagger}AA)^{\ast}]
\end{eqnarray}
and
\begin{eqnarray}
f_H =g'_H H^2 {\rm Tr}(A A^{\dagger})+g_H H_{\mu}(AA^{\dagger})_{\mu \nu}H_{\nu}.
\label{eq:fB}
\end{eqnarray}
In Table~\ref{table-coeffi},
we summarize the coefficients (the GL parameters) calculated in 
the weak coupling limit 
by considering only the excitations around the Fermi surface   \cite{fujita:1972,Richardson:1972xn, sauls-thesis}.
In this limit, $K_1$ and $K_2$ take the same value. 
In the derivation above, 
it should be noted that 
the following relations hold 
\begin{eqnarray}
{\rm Tr}(AA^{\dagger}(AA^{\dagger})^{\ast})&=&{\rm Tr}(AA^T(AA^T)^{\ast}), \nonumber \\
{\rm Tr}((AA^{\dagger})^2(AA^{\dagger})^{\ast})&=&{\rm Tr}((AA^{\dagger})(AA^T)(AA^T)^{\ast}) \nonumber \\ 
{\rm Tr}(AA^T){\rm Tr}(AA^{\dagger}AA^T)^{\ast}&=&{\rm Tr}(AA^T)^{\ast}{\rm Tr}(AA^{\dagger}AA^T)  \nonumber
\end{eqnarray}
because $A_{\mu i}$ is a symmetric tensor.

We ignore the first term with the coefficient $g'_H$ of $f_H$ 
in Eq.~(\ref{eq:fB})
since the effect of this term can be incorporated into  the shift of 
$\alpha$ in $f_{2+4}$ and consequently the phase structure is not modified.

Let us make comments on other systems 
similar to $^3P_2$ superfluids; 
$^3$He superfluids and ultracold atomic gases of spin-2 BEC.
The $^3$He superfluids are also described by a $3 \times 3$ tensor $A_{\mu i}$ 
but it is not  traceless symmetric unlike $^3P_2$ superfluids 
\cite{vollhardt1990superfluid,volovik2009universe}.
The gradient terms are the same in the weak coupling limit. 
Spin-2 BECs are described by Gross-Pitaevskii equation 
of a $3 \times 3$ traceless symmetric tensor \cite{Kawaguchi:2012ii}. 
The sixth order term $f_6$ is absent in the energy functional of 
the Gross-Pitaevskii equation.
In addition, 
the terms proportional to $K_2$ 
in the gradient term $f_{\rm grad}$ 
are absent in spin-2 BECs. 
In other words, the terms proportional to $K_2$ in the gradient term  
exhibits characteristic features of the $^3P_2$ superfluids.

\begin{table*}[t!]
\begingroup
\renewcommand{\arraystretch}{3}
 \begin{tabular}{c|c|c|c|c} 
    $\alpha$ & $K_1=K_2$ & $\beta$ & $\gamma$ & $g_H$  \\ \hline 
    $\displaystyle{\frac{N(0)}{3}\frac{T-T_c}{T}k_F^2}$ & 
    $\displaystyle{\frac{7\zeta (3)}{240M^2}\frac{N(0)}{(\pi T_c)^2}k_F^4}$ & 
    $\displaystyle{\frac{7\zeta (3)}{60}\frac{N(0)}{(\pi T_c)^2}k_F^4}$ &
    $\displaystyle{-\frac{31}{16}\frac{\zeta (5)}{840}\frac{N(0)}{(\pi T_c)^4}k_F^6}$ &    
    $\displaystyle{\frac{7\zeta (3)}{24}\frac{N(0)}{(\pi T_c)^2}\frac{(\gamma_n \hbar)^2}{2(1+F)^2}H^2 k_F^2}$  \\ 
  \end{tabular}
\endgroup
\caption{
The GL parameters in the weak coupling limit. 
In the derivation of $\alpha$, we took $g={3\pi^2 \over M k_F^3}$ 
in order for the $T$ dependence of $\alpha$ to become the same 
with that of the BCS theory.
Here,  $k_F$ is the Fermi momentum defined by $k_F = \hbar c (3\pi^2 \rho)^{1/3}$ where $\rho$ is the neutron density. 
$N(0) \equiv \frac{M k_F}{2\pi^2}$ is the density of states 
$N={M k \over 2\pi^2}$ 
on the Fermi surface {$k=k_F$},
$T_c$ is the critical temperature for the $^3P_2$ superfluidity and
the Riemann zeta function $\zeta (n)$ is defined by $\zeta (n) = \sum_{k=1}^{\infty}\frac{1}{k^n}$,
for which $\zeta (3) \sim 1.202$ 
and $\zeta (5) \sim 1.037$.
$\gamma_n$ is the gyromagnetic ratio of the neutrons
and 
$F$ is the Fermi liquid correction about the Pauli spin susceptibility.
In this paper, we take $F=-0.75$,  $T_c = 0.2$MeV, $T = 0.8 T_c$ and $\rho  = 0.17$/fm$^3$ for numerical simulations. 
}
\label{table-coeffi}
\end{table*}

\subsection{Ground states}
The ground states of the GL free energy with total angular momentum two 
were classified by Mermin \cite{Mermin:1974zz}.
According to this classification, the ground state of 
$^3P_2$ superfluids in the weak coupling limit is in the nematic phase 
\cite{Sauls:1978lna}.

We summarize the effects of each term on the symmetry breaking pattern 
in the cases 
(1) $f_{2+4}$, (2) $f_{2+4}+f_{6}$,  
(3) $f_{2+4} + f_H$, and 
(4) $f_{2+4}+f_{6}+ f_H$.

\bigskip
(1) First, we consider the simplest case $f_{2+4}$. 
The magnitude of the sixth order term is 
much smaller than that of the fourth order term
in the region that the GL theory is appropriate, that is, 
when the gap parameter is small enough. 
Here we consider the energy scale in which 
$f_6$ is negligible.

At the fourth order level, the ground state $A_{4{\rm th}}$ can be written as
\begin{eqnarray}
A_{4{\rm th}}^{(x,y,z)} =
\sqrt{\frac{|\alpha|}{\beta(r^2+(1+r)^2+1)}}
  \left(
    \begin{array}{ccc}
      r & 0 & 0 \\
      0 & -(1+r) & 0 \\
      0 & 0 & 1
    \end{array}
  \right) \nonumber\\
\label{groundstate-4th}  
\end{eqnarray}
with a continuous degeneracy $r$ 
up to the $SO(3)$ action, 
where $(x,y,z)$ implies that we take the Cartesian $xyz$ coordinates for 
the indices of the tensor $A$. 
Here, $r \in \mathbb{R}$ is a parameter 
whose range can be restricted to $-1 \leq r \leq -1/2$ 
without the loss of generality. 
In this range, the eigenvalues in the order parameter have the following magnitude relation
\begin{eqnarray}
(1+r)^2 \leq r^2 \leq 1.
\end{eqnarray}  
The ground states are continuously degenerate 
with parameterized by $r$ \cite{PhysRevLett.97.180412}
and are 
referred as the nematic phase. 
The ground state manifold can be decomposed 
into three region called strata that have 
the isomorphic unbroken symmetries $H$;
the uniaxial nematic (UN) phase for $r=-1/2$,
$D_2$ biaxial nematic ($D_2$ BN) phase for $-1<r<-1/2$, 
and 
$D_4$ biaxial nematic ($D_4$ BN) phase for $r=-1$. 
We summarize the unbroken symmetry $H$, 
the order parameter manifold $G/H$ and 
the homotopy groups from $\pi_0$ to $\pi_4$ of the 
order parameter manifold in Table \ref{table-sym}.

While the order parameter $G/H$ represent for 
gapless Nambu-Goldstone (NG) modes, 
the parameter $r$ here represents for 
an additional gapless mode called a quasi-NG mode, 
that was found in a spin-2 BEC \cite{Uchino:2010pf}.
In the nematic phase, the $SO(3)$ is enhanced to 
$SO(5)$ in the level of the equation of motion,
and when it is spontaneously broken, 
there appears the additional gapless mode, that is, the quasi-NG mode.
The whole ground state manifold 
(extended order parameter manifold) that contains 
both the NG and quasi-NG modes
is ${U(1) \times SO(5) \over {\mathbb Z}_2 \ltimes SO(4)} \simeq 
{U(1) \times S^4 \over {\mathbb Z}_2}$.

\begin{table*}[t!]
\begingroup
\renewcommand{\arraystretch}{3}
 \begin{tabular}{c|c|c|c|ccccc|c} 
    $r$ & Phase & $H$ & $G/H$ &  $\pi_0$ &  $\pi_1$ & $\pi_2$ & $\pi_3$ & $\pi_4$ & Physical situation  \\ \hline 
    $-1/2$ & UN & $O(2)$ &$ U(1) \times [SO(3) / O(2)]$ & $0$ & ${\mathbb Z} \oplus {\mathbb Z}_2$ & ${\mathbb Z}$ & ${\mathbb Z}$ & ${\mathbb Z}_2$ & $f_{2+4}+f_6$\\
    $-1 < r < -1/2$ & $D_2$ BN  & $D_2$ & $ U(1) \times [SO(3) / D_2]$ & $0$ &  ${\mathbb Z} \oplus {\mathbb Q}$ & $0$ & ${\mathbb Z}$ & ${\mathbb Z}_2$ &  $f_{2+4}+f_6+f_H$\\
    $-1$ & $D_4$ BN & $D_4$ & $ [U(1) \times SO(3)] / D_{4}$ & $0$ & ${\mathbb Z} \times_h D_4^*$ & $0$ & ${\mathbb Z}$ & ${\mathbb Z}_2$ & $f_{2+4}+f_H$\\
  \end{tabular}
\endgroup
\caption{The strata in the nematic phase. 
We show the range of $r$, the phase, the unbroken symmetry $H$, 
the order parameter manifold $G/H$, 
the homotopy groups from $\pi_0$ to $\pi_4$, 
and the physical situations (free energy) 
that realize these states.
$*$ indicates the universal covering group,  
and ${\mathbb Q} = D_2^*$ is a quaternion group. 
For the definition of 
the product $\times_h$, see $\S$4.2.2 and Appendix A of Ref.~\cite{Kobayashi:2011xb}.
}
\label{table-sym}
\end{table*}

\bigskip

\begin{figure}
  \begin{center}
   \includegraphics[width=90mm]{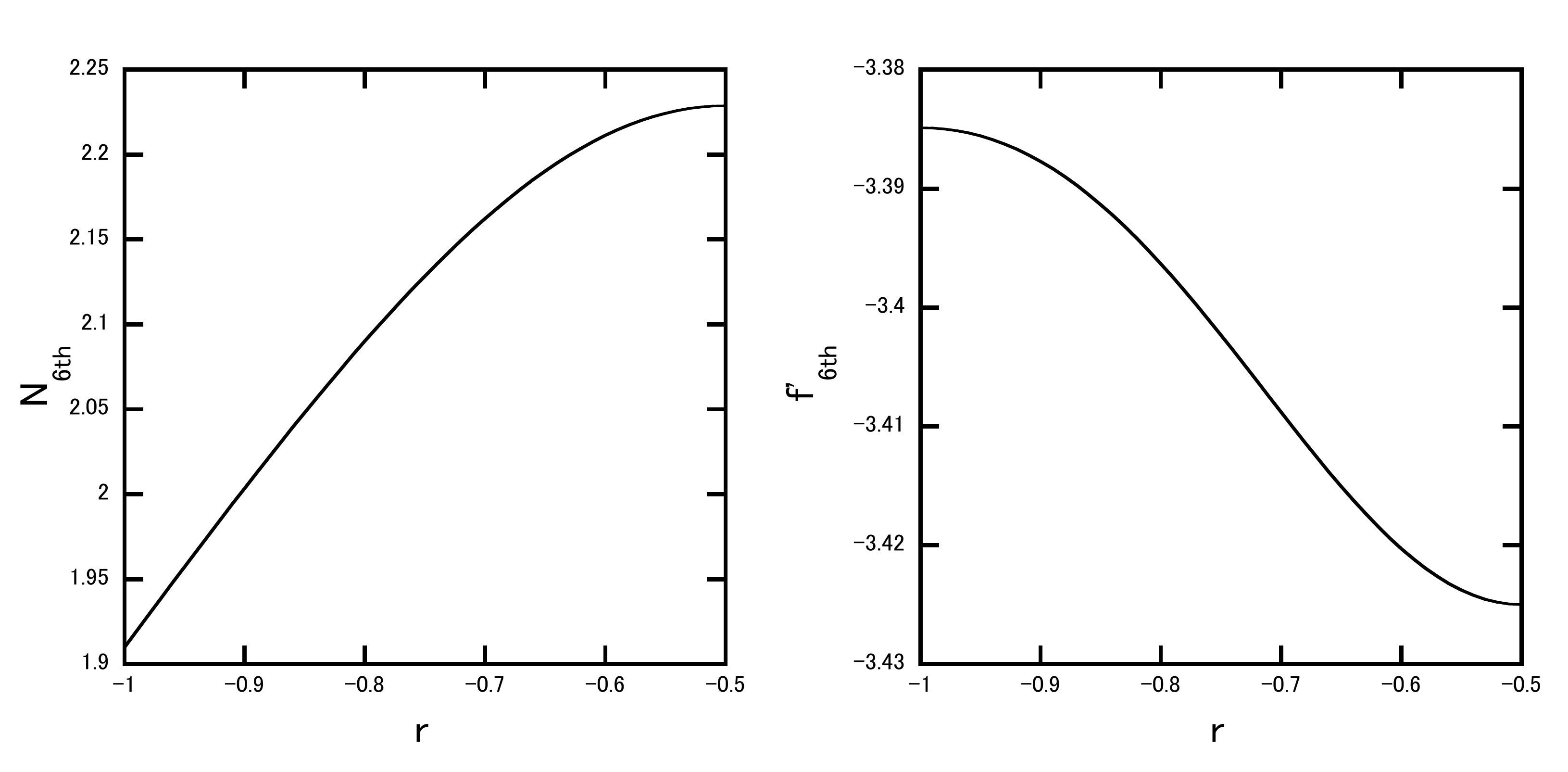}
  \end{center}
  \caption{The normalization and free energy $f_{2+4} + f_6$.
  Left panel: the normalization $N_{6{\rm th}}$ as a function of $r$.
  Right panel: $f'_{{\rm 6th}}$ defined by 
  $f_{2+4} + f_6 \equiv \frac{|\alpha|^2}{6\beta}f'_{{\rm 6th}}$
  as a function of $r$.
  The UN state with $r=-1/2$ is the ground state.
  \label{sixth-free}}
\end{figure}

(2) Next, let us add the sixth order term $f_6$ (so that the total free energy is $f_{2+4}+f_{6}$), 
and see which state is selected by this term. 
In Appendix \ref{sec:ground_state},
we show that the ground state is still in the nematic phase 
 in the presence of the sixth order term, 
with correcting the amplitude of the condensates in 
the previous study \cite{Sauls:1978lna}. 
We can derive the ground state $A_{6{\rm th}}$ exactly by minimizing free energy 
with the Ansatz, 
\begin{eqnarray}
A_{6{\rm th}}^{(x,y,z)} 
= \sqrt{\frac{|\alpha|}{6\beta}} N_{{\rm 6th}}
  \left(
    \begin{array}{ccc}
      r & 0 & 0 \\
      0 & -(1+r) & 0 \\
      0 & 0 & 1
    \end{array}
     \right).
\end{eqnarray}
By minimizing the free energy density
\begin{eqnarray}
f_{2+4}+f_6&=&2\alpha(1+r+r^2)N_{6{\rm th}}^2 \nonumber \\
&+&\beta(2r^4+4r^3+6r^2+4r+2)N_{6{\rm th}}^4  \nonumber \\
&+&\gamma(48r^6+144r^5+312r^4+384r^3 \nonumber \\
&& \quad +312r^2+144r+48)N_{6{\rm th}}^6
\label{6th}
\end{eqnarray}
with respect to $N_{6{\rm th}}$, 
we obtain $N_{6{\rm th}}$ and $f_{2+4}+f_6$.
We plot $N_{6{\rm th}}$ and $f_{2+4}+f_6$ as functions of 
$r$ in Fig.~\ref{sixth-free}.
From this figure,
we find that the case with $r=-1/2$ is realized, 
which is the UN phase.
At $r=-1/2$,
we find $N_{{\rm 6th}}$:
\begin{eqnarray}
N_{{\rm 6th}}= \sqrt{\frac{6\beta-\sqrt{(6\beta)^2-2784\alpha \gamma}}{1392|\gamma|}}.
\end{eqnarray}

In the UN phase, the unbroken symmetry is 
$D_{\infty} \simeq O(2) = SO(2) \rtimes Z_2$, 
where $\rtimes$ denotes a semi-direct product. 

\bigskip

\begin{figure}
  \begin{center}
   \includegraphics[width=90mm]{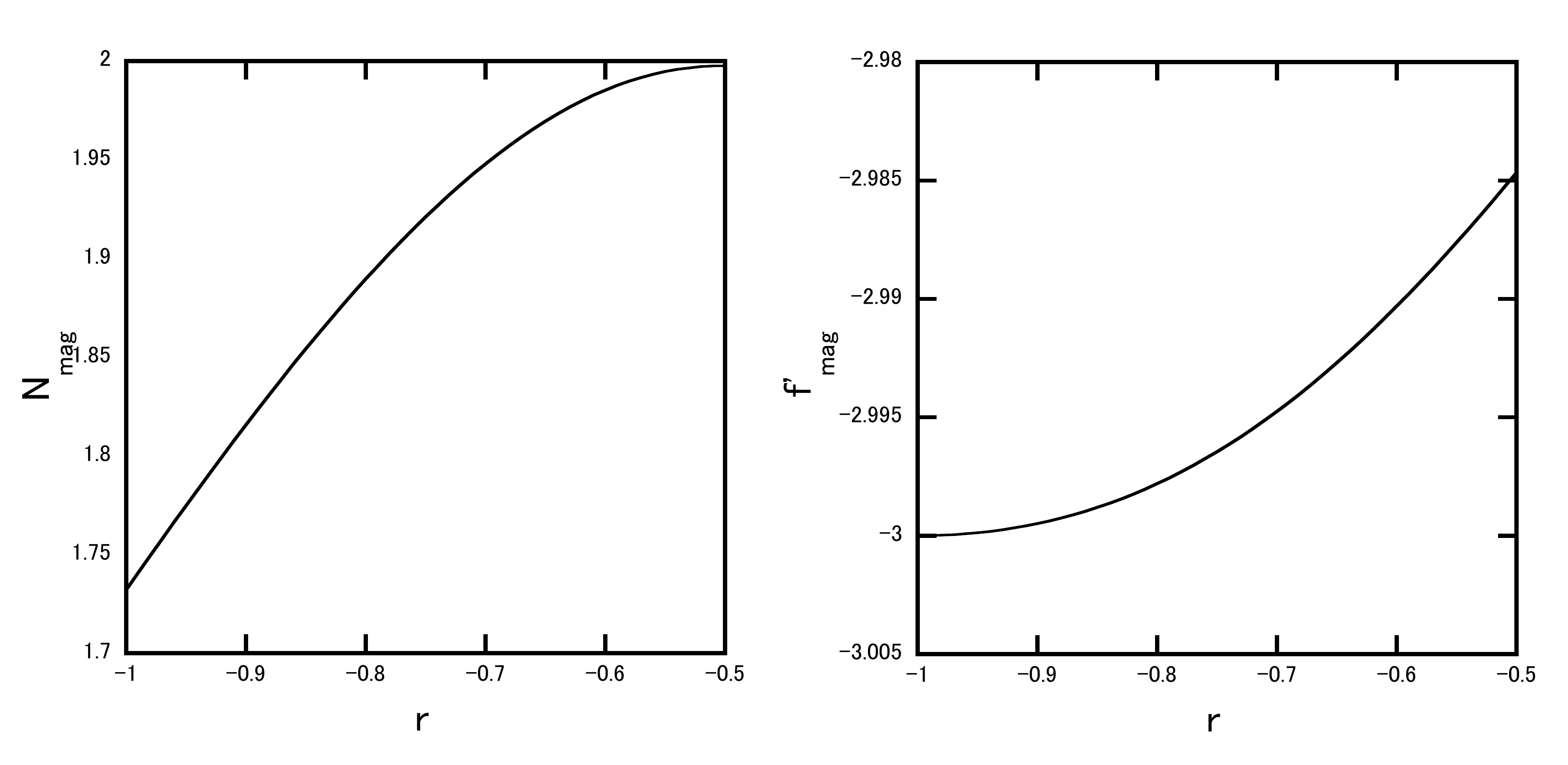}
  \end{center}
  \caption{
The normalization and free energy $f_{2+4} + f_H$.
  Left panel: the normalization $N_{{\rm mag}}$ as a function of $r$.
  Right panel: $f'_{{\rm mag}}$ defined by 
  $f_{2+4} + f_H \equiv \frac{|\alpha|^2}{6\beta}f'_{{\rm mag}}$
  as a function of $r$.
  The $D_4$ BN state  with $r=-1$ is the ground state.
  \label{mag-free}}
\end{figure}

(3) Let us consider the case that we take into account 
the magnetic fields but without the sixth order term:  $f_{2+4}+f_H$.
Here, we consider the two cases for the magnetic fields; 
those along the $z$ (poloidal) direction $(\bm{H} \parallel \bm{z})$
and  the angular (toroidal) direction $(\bm{H} \parallel \bm{\theta})$.
The second case may not be physical for the ground state but 
we use it for the boundary in the presence of a vortex 
in the next section.
We can derive the ground state $A_{{\rm mag}}$ with the same method 
as the case (2).  
Let us consider the magnetic field along the $z$ axis.
With the Ansatz, 
\begin{eqnarray}
A_{{\rm mag}}^{(x,y,z)} 
= \sqrt{\frac{|\alpha|}{6\beta}} N_{{\rm mag}}
  \left(
    \begin{array}{ccc}
      1 & 0 & 0 \\
      0 & r & 0 \\
      0 & 0 & -(1+r)
    \end{array}
     \right),
\end{eqnarray}
we can minimize the free energy density
\begin{eqnarray}
f_{2+4}+f_H 
&=&(2\alpha(1+r+r^2)+g_H H_z^2(1+r)^2)N_{{\rm mag}}^2 \nonumber \\
&+& \beta(2r^4+4r^3+6r^2+4r+2)N_{{\rm mag}}^4  
\label{mag}
\end{eqnarray}
with respect to $N_{{\rm mag}}$.
Fig.~\ref{mag-free} shows
$N_{{\rm mag}}$ and $f_{2+4}+f_H$ 
that minimizing the free energy as functions of $r$.
From this figure,
we find that the case with $r=-1$ is realized, 
which is the $D_4$ BN phase.

To summarize. 
the  ground state $A_{{\rm mag}}$ becomes
\begin{eqnarray} 
\begin{cases}
A^{(x,y,z)}_{{\rm mag}} = 
\displaystyle{\sqrt{\frac{|\alpha|}{2\beta}}}
  \left(
    \begin{array}{ccc}
      1 & 0 & 0 \\
      0 & -1 & 0 \\
      0 & 0 & 0
    \end{array}
  \right)  \ \ \ \ \   (\bm{H} \parallel \bm{z}) \\
A^{(\rho,\theta,z)}_{{\rm mag}} = 
\displaystyle{\sqrt{\frac{|\alpha|}{2\beta}}}
   \left(
    \begin{array}{ccc}
      1 & 0 & 0 \\
      0 &  0 & 0 \\
      0 & 0 & -1
    \end{array}
  \right)  \ \ \ \ \   (\bm{H} \parallel \bm{\theta})
  \end{cases}.
\label{magnetic field} 
\end{eqnarray}
The energy contribution from the magnetic fields vanish 
in these cases.
The symmetry of this state is $D_4$ symmetry, 
and we call this phase as the $D_4$ BN phase \cite{Song:2007ca,Kawaguchi:2012ii} \footnote{
In the context of spinor BEC, this is simply called as the BN phase 
\cite{Song:2007ca,Kawaguchi:2012ii}.
}.

\bigskip 

\begin{figure}
  \begin{center}
   \includegraphics[width=90mm]{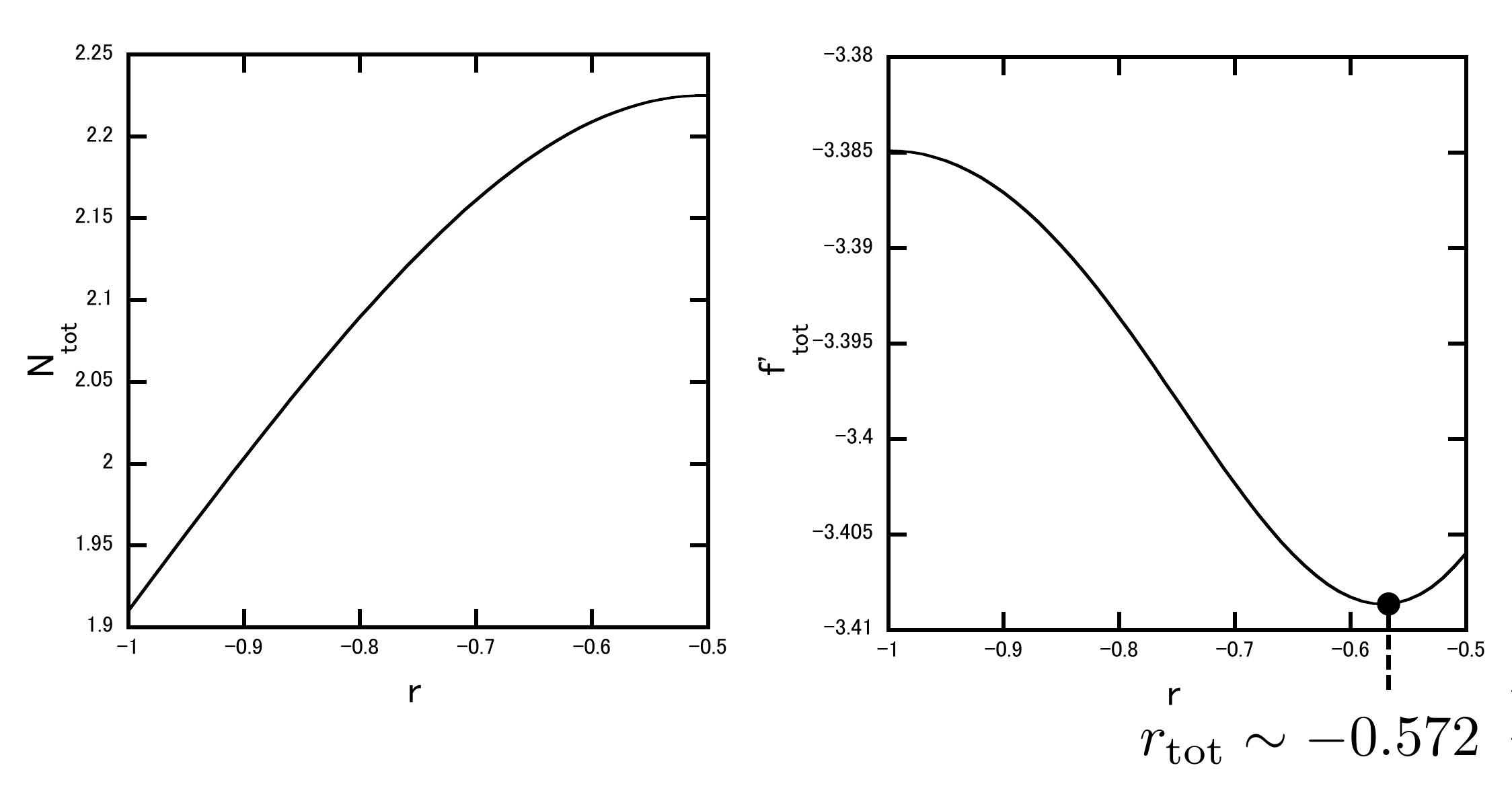}
  \end{center}
  \caption{
The normalization and free energy $f_{2+4} + f_6+f_H$.
  Left panel: the normalization $N_{{\rm tot}}$ as a function of $r$.
  Right panel: $f'_{{\rm tot}}$ defined by 
  $f_{2+4} + f_6+f_H \equiv \frac{|\alpha|^2}{6\beta}f'_{{\rm tot}}$
  as a function of $r$ (for $H=10^{15}$ Gauss).
The intermediate state, the $D_2$ BN state 
(with $r \sim -0.572$ for $H=10^{15}$ Gauss), is the ground state.
  \label{total-free}}
\end{figure}

(4) Finally, we consider the total free energy including 
the sixth order term and magnetic term  ($f_{2+4}+f_{6}+ f_H$),
where we consider the magnetic field along the $z$ axis.  
The intermediate states with the symmetry $D_2$,  
that we call the $D_2$ BN phase, are realized. 
The order parameter has the following form
\begin{eqnarray}
A^{(x,y,z)}= N_{{\rm tot}}
\left(
    \begin{array}{ccc}
      1 & 0 & 0 \\
      0 & r & 0 \\
      0 & 0 & -1-r
    \end{array}
  \right).  
\end{eqnarray}
The free energy density $f_4+f_6+f_H$ can be written in terms of 
the parameter $r$ and $N_{{\rm tot}}$ as follows:
\begin{eqnarray}
 f_{2+4}+f_6+f_H 
 &=&(2\alpha(1+r+r^2)+g_H H_z^2(1+r)^2)N_{{\rm tot}}^2 \nonumber \\
&+&\beta(2r^4+4r^3+6r^2+4r+2)N_{{\rm tot}}^4  \nonumber \\
&+&\gamma(48r^6+144r^5+312r^4+384r^3 \nonumber \\
&& +312r^2+144r+48)N_{{\rm tot}}^6.
\label{6th+mag}
\end{eqnarray}
By minimizing this free energy density with respect to $r$ and $N_{{\rm tot}}$, we can obtain the ground state.
In Fig. \ref{total-free},
we plot $N_{{\rm tot}}$ and the free energy density as a function of $r$ with $H=10^{15}$ Gauss.
In this case,
the minimum free energy density can be achieved at 
\begin{eqnarray}
 r \sim -0.572 \equiv r_{{\rm tot}} 
\mbox{ for }  H=10^{15} \mbox{ Gauss}.  \label{eq:rtot}
\end{eqnarray}

\begin{figure}
  \begin{center}
   \includegraphics[width=80mm]{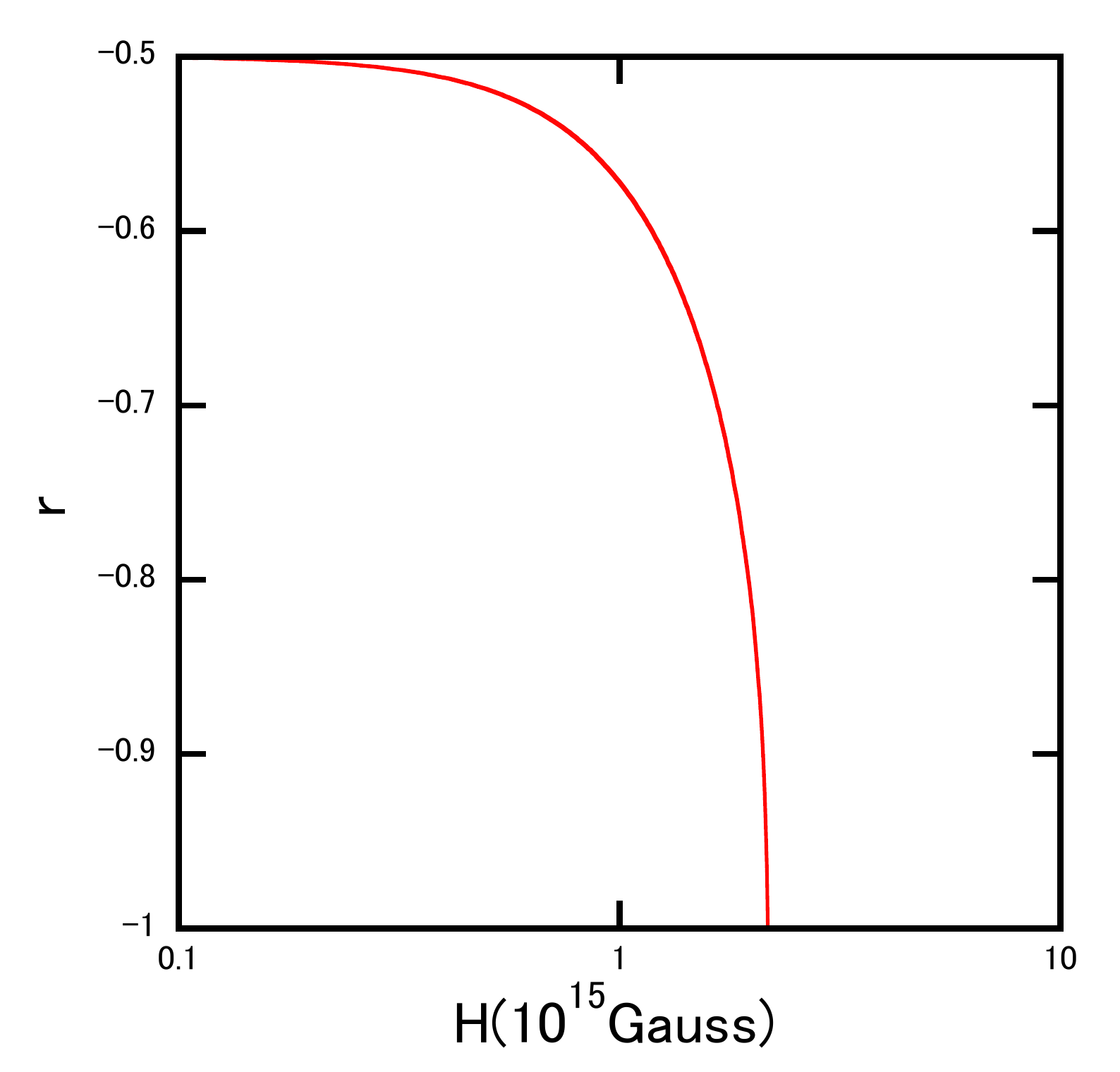}
  \end{center}
  \caption{The parameter $r = r_{\rm tot}(H)$ minimizing the free energy 
$f_{2+4}+f_6+f_H$ as a function of magnetic field $H$.
  \label{condition-f}}
\end{figure}

In Fig.~\ref{condition-f},
we plot $r$ that minimizing the energy 
as a function of the strength of the magnetic field $H$.
The ground state is in the UN phase
for the weak magnetic field, 
while the $D_4$ BN phase is realized 
for the strong magnetic field.
The parameter $r$ changes drastically 
around the magnetic field of $10^{15}$ Gauss.

\section{Vortex structures 
in $^3P_2$ superfluids} \label{sec:vortex}

\subsection{Vortex lattice}
When superfluids are rotating, superfluid vortices are created along 
the rotation axis. 
In this section, we discuss vortices in the $^3P_2$ superfluids.
The existence of vortices in the $^3P_2$ superfluids is topologically ensured by 
the first homotopy group summarized in Table \ref{table-sym}.  
The number $N_v$ of vortices with the unit circulation 
created inside rotating neutron stars can be estimated to be
\begin{eqnarray}
N_v \sim 1.9\times 10^{19} \left(\frac{1{\rm ms}}{P}\right)\left(\frac{M^*}{900{\rm MeV}}\right)\left(\frac{R}{10{\rm km}}\right)^2 \label{eq:Nv}
\end{eqnarray} 
where $P$ is the period of the neutron star, 
$M^*$ is the effective neutron mass, 
and $R$ is the radius of the ${}^3P_2$ superfluid.
Then, we can estimate the distance between vortices $d$ from
\begin{eqnarray}
\pi d^2 \times N_v =\pi R^2,
\end{eqnarray}
that implies the intervortex distance $d$ to be
\begin{eqnarray}
d \sim 1.7 \times 10^{-6} {\rm m} 
\end{eqnarray}
for the typical values for $P$, $M^*$ and $R$ in Eq.~(\ref{eq:Nv}).

On the other hand, the coherence length $\xi$ of $^3P_2$ superfluid is about $10-100$ fm.
Therefore the distance between vortices is much larger than the coherence length,
and therefore we consider a single vortex below. 

\subsection{Vortex Ansatz and 
asymptotic energy of a vortex}
Let us derive the equation of motion from the free energy $F$ introduced in the last section.
Here, we consider the following Ansatz for the order parameter of a vortex state:  
\begin{eqnarray}
&& A^{(x,y,z)} 
= \sqrt{\frac{|\alpha|}{6\beta}}R(n\theta)A^{(\rho, n\theta, z)}R^T(n\theta)e^{i l \theta}, \nonumber \\
&& A^{(\rho, n\theta, z)} =
  \left(
    \begin{array}{ccc}
      f_1 & ige^{im\theta +i\delta} & 0 \\
      ige^{im\theta+ i\delta} & f_2 & 0 \\
      0 & 0 & -f_1-f_2
    \end{array}
  \right)    \label{eq:ansatz}
\end{eqnarray}
in the cylindrical coordinates $(\rho,\theta,z)$, where 
$l,m,n$ are integers, $l,m,n \in {\mathbb Z}$, explained below, 
$\delta$ is a constant,  
$A^{(x,y,z)}$ is the order parameters in the Cartesian basis and 
$A^{(\rho, n\theta, z)}$ is the order parameters in 
the cylindrical basis ($n=1$) or its higher generalizations, 
which are related by 
a rotation matrix  $R$, given by 
\begin{eqnarray}
R(n\theta)=
\left(
    \begin{array}{ccc}
      {\rm{cos}}n\theta & -{\rm{sin}}n\theta & 0 \\
      {\rm{sin}}n\theta & {\rm{cos}}n\theta & 0 \\
      0 & 0 & 1
    \end{array}
  \right) .
\end{eqnarray} 
In Eq.~(\ref{eq:ansatz}), $f_1,f_2,g$ are profile functions depending only 
on the radial coordinate $\rho$, 
and 
the boundary conditions for them are 
\begin{eqnarray}
 && f_1,f_2 \to  \mbox{constant}, \quad  g \to 0  
 \quad\quad\quad \;\;\mbox{ as } 
\rho \to \infty,   \nonumber \\
 && f_1,f_2 \to 0, \quad 
 \left\{\begin{array}{c}
   g \to 0 \mbox{ for } n \neq -1  \\
   g' \to 0 \mbox{ for } n=-1
 \end{array} \right.
 \mbox{ as } \rho \to 0,
\end{eqnarray}
where the case of $n = -1$ is exceptional since the 
the total winding of $g$ vanishes.
As denoted,
the configuration is labeled by the three integers 
$l,m,n \in {\mathbb Z}$, where 
$l$ is the winding number of the vortex,  
$n$ represents a rotation of $SO(3)$ that does not have 
a topological nature, 
and 
$m$ is a semi-topological winding number (relative to that of $f_{1,2}$) 
defined locally for the component $g$ \footnote{ 
Since the boundary condition for $g$ are zero for $\rho=0,\infty$ (for $n \neq -1$), 
it is a ring-shape if it appears, and $m$ denotes how many times
the phase of $g$ is twisted along the ring,  
see Ref.~\cite{Kobayashi:2012ib} for a similar example in a spinor BEC.   
}.

Let us consider the tension, the energy per unit length, of the vortex. 
A bulk part of free energy density, $f_{2+4}+f_6+f_H$, 
does not depend on the integers $l,m,n$, 
while the gradient term depends on them; 
The leading contribution to the gradient energy at large $\rho$ depends on $n$ and $l$ as follows:
\begin{eqnarray} \label{log-term}
F &=&\int d^2 \rho \frac{1}{\rho^2} 
\big[
2K_1 (l^2(f_1^2+f_1f_2+f_2^2) + n^2(f_1-f_2)^2) 
\nonumber \\
&&+ 
2K_2(
f_1^2{\rm sin}^2(n-1)\theta+f_2^2{\rm cos}^2(n-1)\theta) 
\nonumber \\ && 
+n^2(f_1-f_2)^2\big] \nonumber \\
&\sim&
2\pi{\rm log} L \nonumber \\
& \times & 
   [ 2K_1 (l^2(f_1^2+f_1f_2+f_2^2)   +n^2(f_1-f_2)^2) ) \nonumber \\
&&+
\begin{cases}
2K_2(l^2f_2^2+n^2(f_1-f_2)^2) ] \ \ \ \ \ \ \ \ \ \ \ \ \   (n=1) \\
K_2\left(l^2(f_1^2+f_2^2)+2n^2(f_1-f_2)^2\right) ]  \ \ \    (n \neq 1) \;\;
\end{cases}
\label{eq:vortex-asym}
\end{eqnarray}
where $\sim$ denotes the asymptotic form and $L$ is the system size 
transverse to the vortex, and 
$f_1$ and $f_2$ in the last line are the boundary values evaluated 
at $\rho \to \infty$. 

\begin{figure}
  \begin{center}
   \includegraphics[width=87mm]{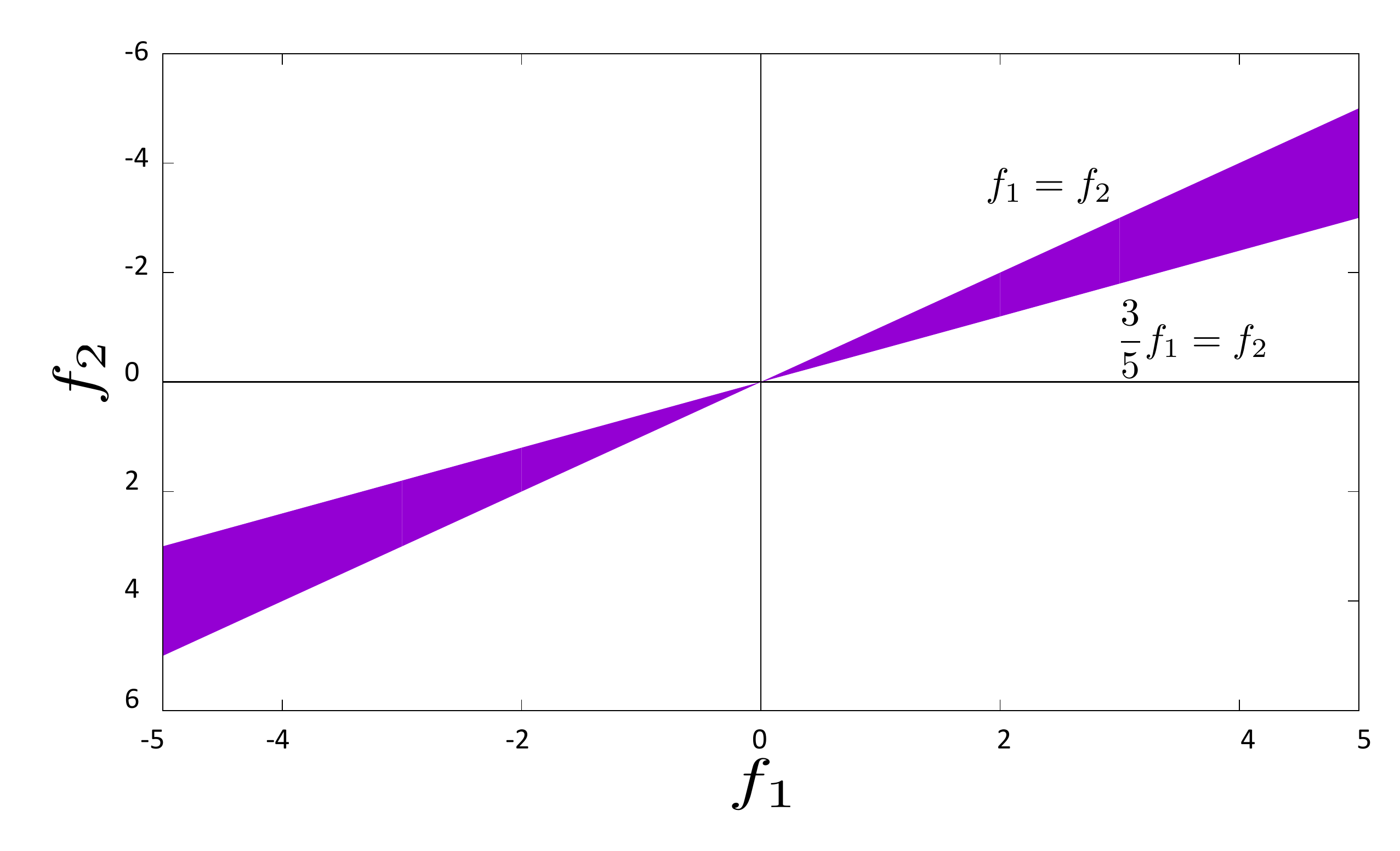}
  \end{center}
  \caption{The conditions for determining whether the Cartesian basis or cylindrical basis is realized.
  The cylindrical basis $(n=1)$ is realized in the shaded area 
while the Cartesian basis $(n=0)$ is realized in the rest.}
  \label{condition-f2}
\end{figure}

From this equation, 
we first see that the configuration with $l=1$ has 
the lower energy than the configuration with higher winding numbers $l>1$,
as for conventional superfluids,
thereby
a vortex with the higher winding $l$ is unstable  
to be split into $l$ unit winding vortices.
In the following, we concentrate on $l=1$.

As for $n$, we find that 
either the case of $n=0$ (the $xyz$-basis) or of $n=1$ (the cylindrical basis) 
gives the lowest free energy. 
The condition on $f_1$ and $f_2$ that determines which configuration 
with $n=0$ or $n=1$ has lower energy 
is plot in Fig.~\ref{condition-f2}, 
where 
the cylindrical basis $(n=1)$ gives lower energy in the shaded region defined by 
\begin{eqnarray}
 {3\over 5}f_1<f_2<f_1,
\end{eqnarray}
while the $xyz$ basis $(n=0)$ gives lower energy in the rest.

This condition can also be translated in terms of the parameter $r$.
If $f_2^2 \leq f_1^2 \leq (f_1-f_2)^2$ is not satisfied, 
the configuration with $n=0$ always give the least energy state.
Therefore, if $n=1$ is realized, the order parameter becomes the following form
\begin{eqnarray}
&& A^{(x,y,z)}   \nonumber \\ 
&& \propto
\sqrt{\frac{1}{r^2+r+1}}
R(n\theta)
  \left(
    \begin{array}{ccc}
      r & 0 & 0 \\
      0 & -1-r & 0 \\
      0 & 0 & 1
    \end{array}
  \right) 
R^T(n\theta)
  e^{i \theta}. \;\;
\end{eqnarray}

\begin{figure}
  \begin{center}
   \includegraphics[width=90mm]{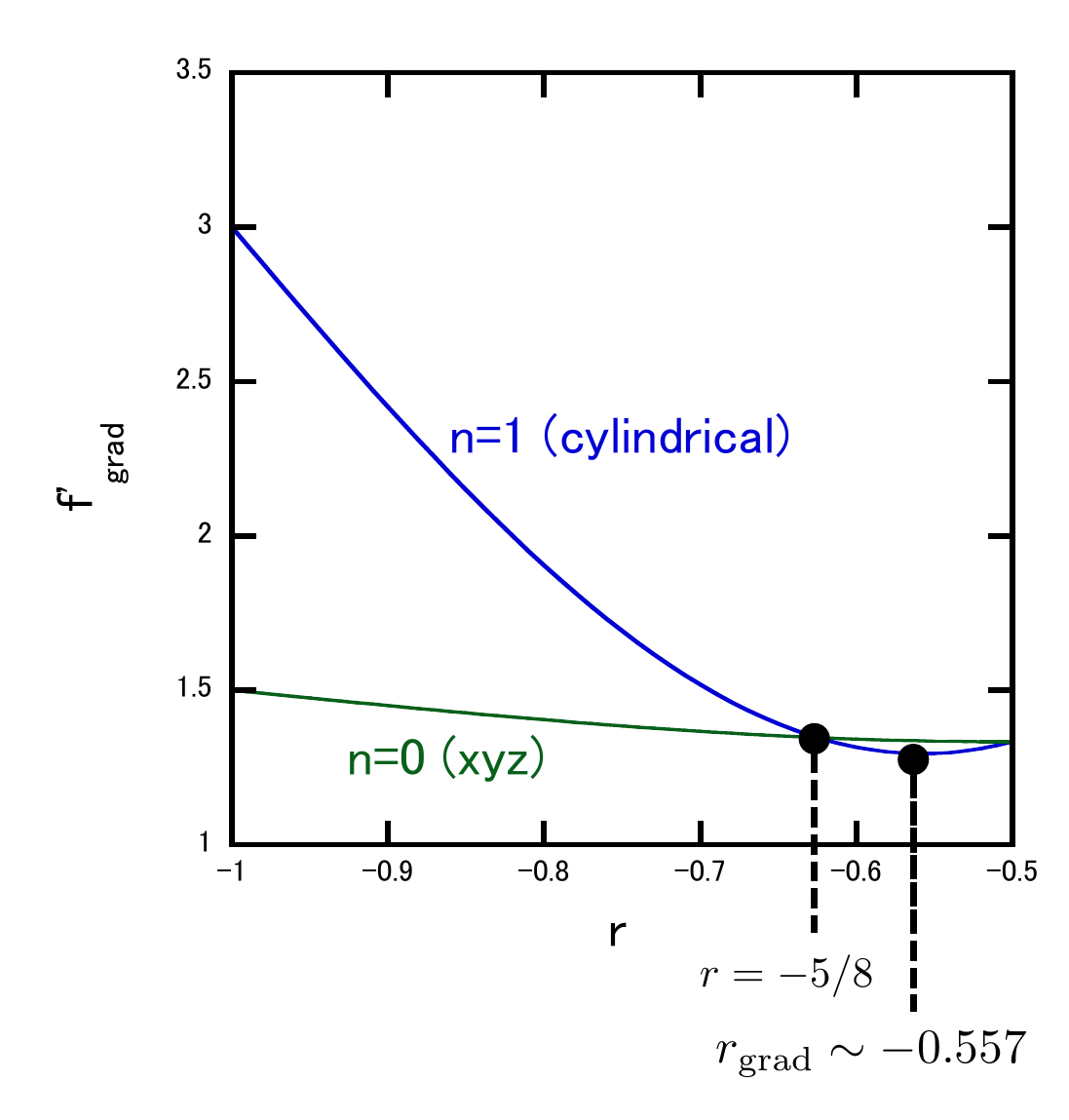}
  \end{center}
  \caption{
  The leading contribution to the free energy 
$F \sim 2\pi K_1{\rm log}L \times f'_{{\rm grad}}$ in Eq.~(\ref{eq:vortex-asym})
as functions of the parameter $r$ at the infinity  
for the $n=0$ (purple curve) and $n=1$ (green curve)  cases, 
which cross at $r = -5/8$.
The minimum point is $r_{\rm grad} = 6 - \sqrt{43}$ 
for $n=1$ in Eq.~(\ref{eq:rgrad}).
}
  \label{condition-f-2}
\end{figure}

In Fig.~\ref{condition-f-2}, 
we plot the leading contribution to  
the free energy proportional to ${\rm log}L$
in Eq.~(\ref{eq:vortex-asym})
as a function of $r$ for the cases of $n=0$ (purple curve) and $n=1$ (green curve).
From this figure, we can see that the configuration of 
$n=1$ has the lower energy in the region $-5/8 \leq r \leq -1/2$.
At the cross point $r=-5/8$, there is a first order phase transition.
In the following, we consider only 
the cases of $n=0$ and $n=1$.

\begin{table*}[ht!]
  \begin{tabular}{c|c|c|c|c|c} \hline
    & Case (1) & Case (2) & Case (3a) & Case (3b) & Case (4) \\ \hline \hline
    Phase &$D_2$ BN & UN & $D_4$ BN & $D_4$ BN & 
$D_2$ BN (or UN, $D_4$ BN)\\  \hline
    $\left(
    \begin{array}{ccc}
      f_1 &  &  \\
       & f_2  &  \\
       &  & -f_1-f_2
    \end{array}
  \right)$ at $\rho \rightarrow \infty$  & 
  $\left(
    \begin{array}{ccc}
       r_{{\rm grad}} &  &  \\
      & -1-r_{{\rm grad}} &  \\
      &  & 1
    \end{array}
  \right)$ &
    $\left(
    \begin{array}{ccc}
      1 &  &  \\
       & 1  &  \\
       &  & -2
    \end{array}
  \right)$ 
   &
     $\left(
    \begin{array}{ccc}
      1 &  &  \\
       & 0  &  \\
       &  & -1
    \end{array}
  \right)$ 
   &
     $\left(
    \begin{array}{ccc}
      1 &  &  \\
       & -1  &  \\
       &  & 0
    \end{array}
  \right)$ 
   & 
    $\left(
    \begin{array}{ccc}
      1 &  &  \\
       & r_{{\rm tot}}  &  \\
       &  & -1-r_{{\rm tot}}
    \end{array}
  \right)$ 
   \\   \hline
   Basis & $n=1$ & $n=0$ (or $1$) & $n=1$ & $n=0$ & $n=0$ \\ \hline
    Situation ($f_{\rm grad}+f_{2+4}$ plus) & non  & $f_6$ & 
 $f_{H(||\bm{\theta})}$  & $f_{H(||\bm{z})}$ 
& $f_6+f_{H(||\bm{z})}$  \\ \hline
  \end{tabular}
\caption{The boundary conditions and physical situation for each case. 
 $ r_{\rm grad} \equiv 6-\sqrt{43} \sim -0.557$ given in Eq.~(\ref{eq:rgrad}), 
and $r_{\rm tot}$ depends on the magnetic field 
($r_{\rm tot} \sim -0.572$ in Eq.~(\ref{eq:rtot}) for $H=10^{15}$ Gauss).
}
\label{table-case}
\end{table*}

\subsection{Tension of a vortex}
We calculate the free energy per the unit vortex length analytically for each basis:
\begin{eqnarray}
F&=&\int d^2 \rho \ \frac{|\alpha|}{6\beta}
\big( K_1 t_1+K_{2} t_2+\alpha t_3  
+\frac{|\alpha|}{6\beta}\beta t_4+\frac{\alpha^2}{36\beta^2} \gamma t_5 \big)
\nonumber \\
\end{eqnarray}
where 
the terms $t_1$ and $t_2$ come from 
the gradient terms and 
the terms $t_3$, $t_4$ and $t_5$ come from
the second, fourth and sixth order terms, respectively, 
in the GL free energy density in Eq.~(\ref{freeenergydensity}).
Here $t_{1,2}$ can be written in the cylindrical basis as
\begin{eqnarray}
t^{(\rho,\theta,z)}_1&=&2(f_1'^2+f_2'^2+f_1'f_2'+g'^2) \nonumber \\
&+& \frac{1}{\rho^2}(4f_1^2+4f_2^2-2f_1f_2+(8+2(m+1)^2)g^2 \nonumber \\
&-& 4(f_1-f_2)g(m+2){\rm cos}(m\theta +\delta)), \\
\label{gradient-1}
t^{(\rho,\theta,z)}_2&=&2(f_1'^2+g^2) \nonumber \\
&+&\frac{2}{\rho^2}(f_2^2+4g^2+(m+1)^2g^2+(f_1-f_2)^2 \nonumber \\
&-&2g(m+1)(f_1-f_2){\rm cos}(m\theta +\delta) \nonumber \\
&+&4f_2g{\rm cos}(m\theta +\delta)) \nonumber \\
&+&\frac{1}{\rho}(-2(f_1'+f_2')g(m+1){\rm cos}(m\theta +\delta) \nonumber \\
&+&2(f_1+f_2)g'{\rm cos}(m\theta +\delta) \nonumber \\
&+&2(f_1'+f_2')(f_1-f_2)), 
\label{gradient-2}
\end{eqnarray}
and in the $xyz$ basis as 
\begin{eqnarray}
t^{(x,y,z)}_1&=&2(f_1'^2+f_2'^2+f_1'f_2'+g'^2) \nonumber \\
&+&\frac{2}{\rho^2}(f_1^2+f_2^2+f_1f_2+(m+1)^2g^2), \\
t^{(x,y,z)}_2&=&2({\rm cos}^2\theta f_1'^2+{\rm sin}^2\theta f_2'^2+g'^2) \nonumber \\
&+&\frac{2}{\rho^2}({\rm sin}^2 \theta f_1^2+{\rm cos}^2 \theta f_2^2+(m+1)^2g^2) \nonumber \\
&-&2{\rm sin}2\theta{\rm sin}(m\theta +\delta)(f_1'+f_2')g' \nonumber \\
&-&\frac{2}{\rho}{\rm cos}2\theta{\rm cos}(m\theta +\delta)(m+1)(f_1'+f_2')g \nonumber \\
&+&\frac{2}{\rho}{\rm cos}2\theta{\rm cos}(m\theta +\delta)(f_1+f_2)g' \nonumber \\
&+&\frac{2}{\rho^2}{\rm sin}2\theta{\rm sin}(m\theta +\delta)(m+1)(f_1+f_2)g.
\end{eqnarray} 
The rest terms of the free energy density
$t_{3,4,5}$ can be written in the both basis as 
\begin{eqnarray}
t_3=&&2(f_1^2+f_2^2+f_1f_2+g^2), 
\end{eqnarray}
\begin{eqnarray}
t_4=&&2f_1^4+4f_1^3f_2+6f_1^2f_2^2+4f_1+f_2^3+2f_2^4  \nonumber \\
&&+((6+2{\rm cos}2(m\theta +\delta))f_1^2 
+4f_1f_2 \nonumber \\
&&+(6+2{\rm cos}2(m\theta +\delta))f_2^2)g^2+2g^4,
\end{eqnarray}
\begin{eqnarray}
t_5&=&48f_1^6+144f_1^5f_2+312f_1^4f_2^2+384f_1^3f_2^3 \nonumber \\
&+&312f_1^2f_2^4+144f_1f_2^5+48f_2^6+48g^6  \nonumber \\
&+&((288+144{\rm cos}2(m\theta +\delta))f_1^4 \nonumber \\
&+&(360+120{\rm cos}2(m\theta +\delta))f_1^3f_2 \nonumber \\
&+&(576+240{\rm cos}2(m\theta +\delta))f_1^2f_2^2 \nonumber \\
&+&(360+120{\rm cos}2(m\theta +\delta))f_1f_2^3 \nonumber \\
&+&(288+144{\rm cos}2(m\theta +\delta))f_2^4)g^2 \nonumber \\
&+&((288+120{\rm cos}2(m\theta +\delta))f_1^2\nonumber \\
&+&(144-48{\rm cos}2(m\theta +\delta))f_1f_2 \nonumber \\
&+&(288+120{\rm cos}2(m\theta +\delta))f_2^2)g^4.
\end{eqnarray}

The effect of $\delta$ on the free energy density is not clear.
In this paper, we restrict the case with $\delta=0$, 
which is consistent with the equation of motion, 
since the imaginary part of non-diagonal elements is directly connected with real part of diagonal elements
through the equation of motion.
By substituting the order parameter into Eq.~(\ref{freeenergydensity}) 
and differentiating it with respect to $f_1,\ f_2$ and $g$, 
we obtain the sets of the equation of motions for each basis,
as summarized in Appendix \ref{sec:eom}.

\subsection{Vortex solutions}

In this paper, we construct the vortex configurations in
 the following five cases: 
\begin{itemize}
\item Case (1): $F=\int d^2\rho\ f_4$
\item Case (2): $F=\int d^2\rho\ (f_4+f_6)$
\item Case (3a): $F=\int d^2\rho\ (f_4+f_H)$ and $\bm{H} \parallel \bm{\theta}$. 
\item Case (3b): $F=\int d^2\rho\ (f_4+f_H)$ and $\bm{H} \parallel \bm{z}$. 
\item Case (4): $F=\int d^2\rho\ (f_4+f_6+f_H)$ and $\bm{H} \parallel \bm{z}$.   
\end{itemize}
We summarize the forms of the order parameters in these cases in Table~\ref{table-case}. 
In numerical simulations,
we change the variable $\rho$ ($0 \leq \rho < \infty$) by 
tanh$\rho$ ($0 \leq {\rm tanh}\rho < 1$).
We divide the domain of tanh$\rho$ into 100 parts
and solve the equations of motion given in Appendix \ref{sec:eom} simultaneously
by using the Newton's method.

\bigskip
\underline{Case (1)}

This is only the case studied before 
\cite{Richardson:1972xn,Muzikar:1980as,Sauls:1982ie}. 
This case may be thought to be unphysical 
because of the presence of the sixth order term,
but it is relevant when the gap is small enough 
($f_6$ is much smaller than $f_4$)
and the magnetic field is small compared with $f_4$.
In this case, 
the ground state takes the form in Eq.~(\ref{groundstate-4th}) 
due to the fourth order term. 
We can see from Eq.~(\ref{log-term}) that
the leading part of free energy proportional to log$L$ becomes lower when $f_2^2 \leq f_1^2$,
so that we take the following order parameter form
\begin{eqnarray}
&& A^{(x,y,z)}  \nonumber \\
&& \propto \sqrt{\frac{1}{r^2+r+1}}
R(n\theta)
  \left(
    \begin{array}{ccc}
      r & 0 & 0 \\
      0 & -1-r & 0 \\
      0 & 0 & 1
    \end{array}
  \right) 
 R^T(n\theta)
  e^{i \theta} \quad
\end{eqnarray} 
with $-1 \leq r \leq -1/2$.
Let us assume that the ground state is in the cylindrical basis ($n=1$) from Fig.~\ref{condition-f-2}.
By putting the order parameter into
Eqs.~(\ref{gradient-1}) and (\ref{gradient-2}),
we obtain
\begin{eqnarray}
F \propto \frac{9r^2+10r+3}{r^2+r+1} {\rm log}L.
\end{eqnarray} 
Then, by differentiating this free energy with respect to $r$,
we get \cite{Richardson:1972xn,Muzikar:1980as,Sauls:1982ie} 
\begin{eqnarray}
 r=6-\sqrt{43} \sim -0.557 \equiv r_{\rm grad}  \label{eq:rgrad}
\end{eqnarray} 
that satisfies $-5/8<r_{\rm grad}<-1/2$, and so the lowest energy 
state is in the cylindrical basis ($n=1$) consistently.
It is interesting to emphasize that the particular $r$ is selected 
as the vortex boundary state although 
the ground states 
are continuously degenerate with $r$.

Let us make a comment on the case of  spin-2 BECs 
for which the gradient term consists of only the first term proportional to 
$K_1$ with $K_2=0$.
In this case, 
the boundary becomes the UN phase with $r=-1/2$ (up to the fourth order).
This phenomenon of selection of the vortex boundary state 
is not known  in the context of BEC. 
The quasi-NG mode is gapped through the gradient term 
in the presence of a vortex.

\begin{figure*}[htbp]
  \begin{center}
    \begin{tabular}{c}

      \begin{minipage}{0.33\hsize}
        \begin{center}
          \includegraphics[clip, width=6cm]{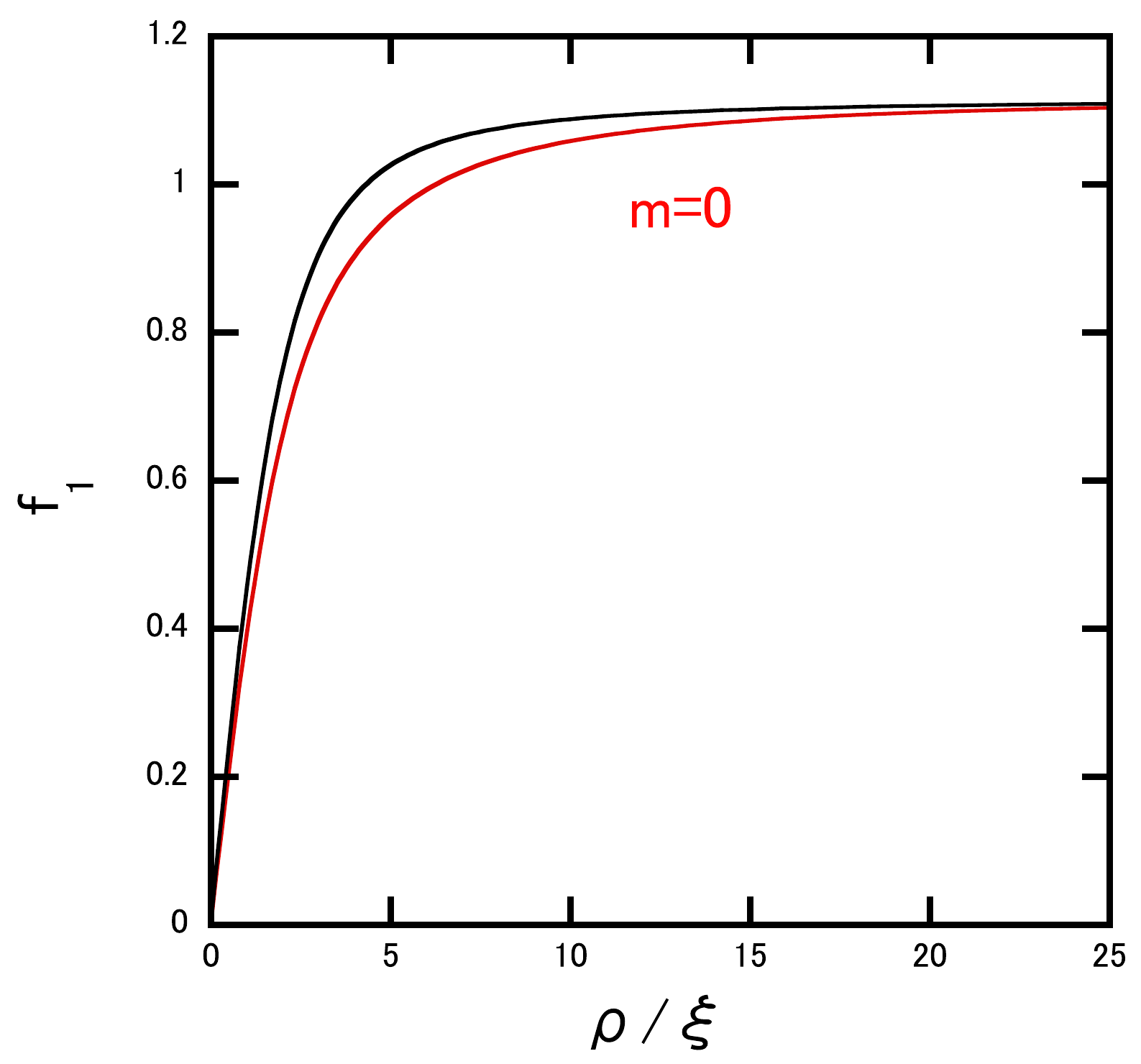}
        \end{center}
      \end{minipage}
      
      \begin{minipage}{0.33\hsize}
        \begin{center}
          \includegraphics[clip, width=6cm]{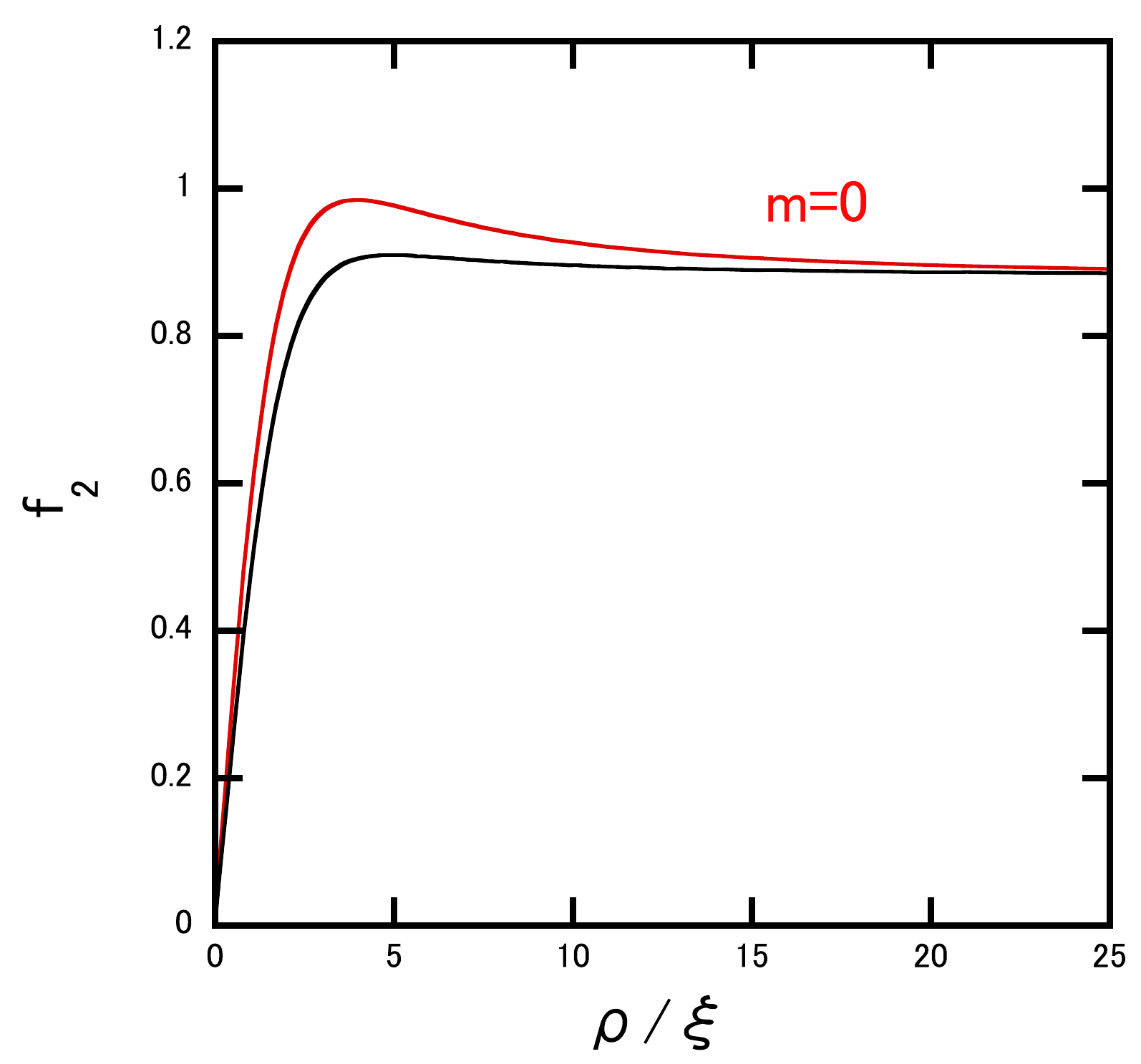}
        \end{center}
      \end{minipage}

      \begin{minipage}{0.33\hsize}
        \begin{center}
          \includegraphics[clip, width=6cm]{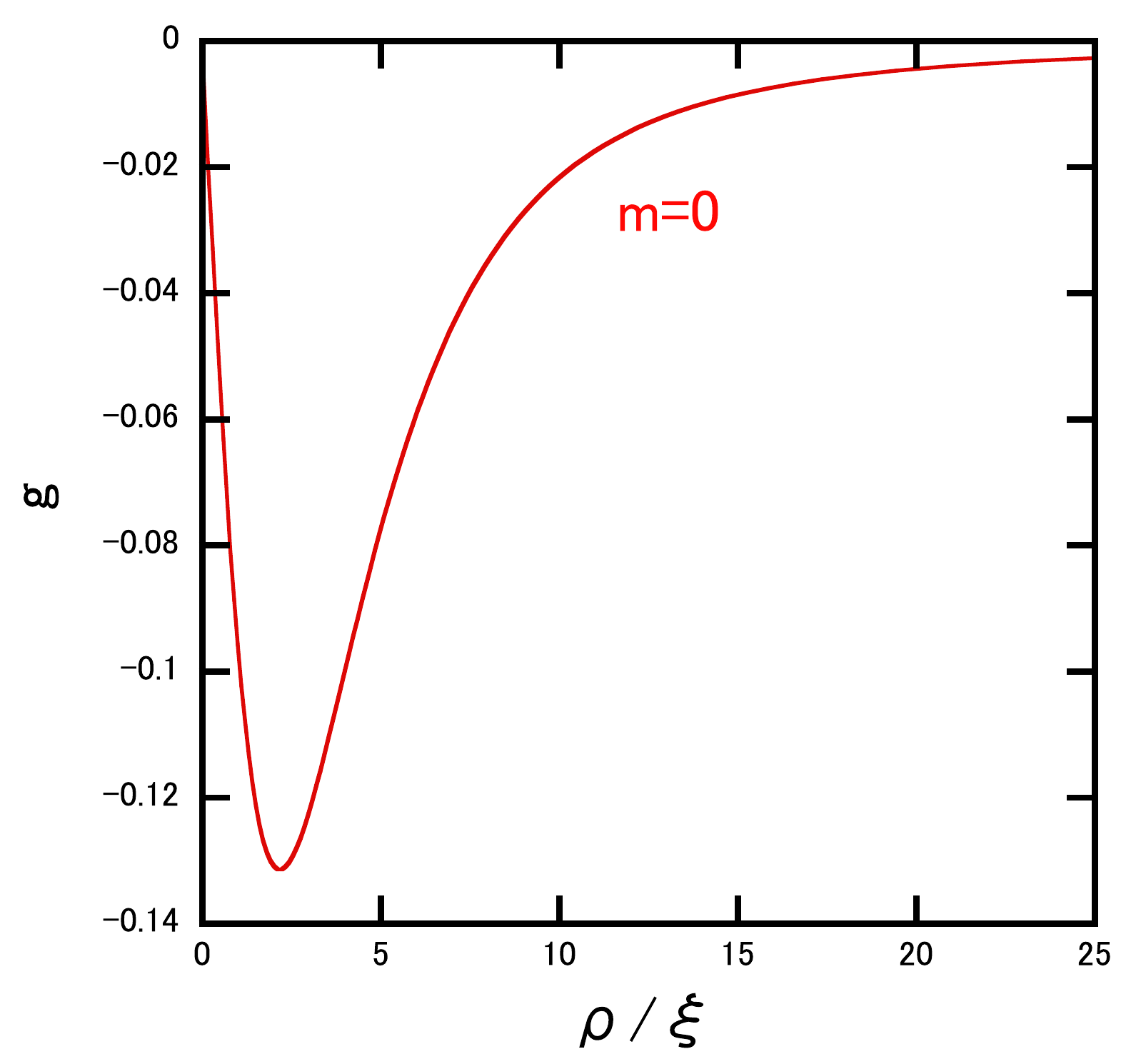}
        \end{center}
      \end{minipage}

    \end{tabular}
    \caption{The profile functions $f_1$, $f_2$ and $g$ as functions of the distance $\rho/\xi$ 
from the vortex center in the case (1).
    The red and black curves correspond to the cases of 
$g \neq 0$ with $m=0$ and $g = 0$, respectively.}
    \label{fig:case1}
  \end{center}
\end{figure*}

In Fig.~\ref{fig:case1}, we plot the profiles 
functions $f_1$, $f_2$ and $g$ as functions of 
the distance $\rho$ from the vortex center. 
The red curves correspond to the case of $m=0$ with $g \neq 0$,
which is the case considered in Ref.~\cite{Sauls:1982ie} 
without explicit solutions.
The new solution here is  the case with $g = 0$ 
represented by the black curves. 
We see that all the terms in the equation of motion for 
$g$ 
[Eq.~(\ref{eom-cyl}) in Appendix \ref{sec:eom}]  
contain $g$ if $m \neq 0$.
Consequently, in this case, 
$g = 0$ is a trivial solution in the entire region of $\rho$ 
since $g$ is fixed to be zero at the both boundaries. 
This implies that all $m \neq 0$ give the same solution 
of $g=0$.
In order to determine which state 
between $g \neq 0$ (with $m=0$) 
and $g=0$ 
has less energy,
we should compare the free energies for these two cases, 
but we leave it as a future problem.
 
\bigskip
\underline{Case (2)}

\begin{figure*}[htbp]
  \begin{center}
    \begin{tabular}{c}

      \begin{minipage}{0.33\hsize}
        \begin{center}
          \includegraphics[clip, width=6cm]{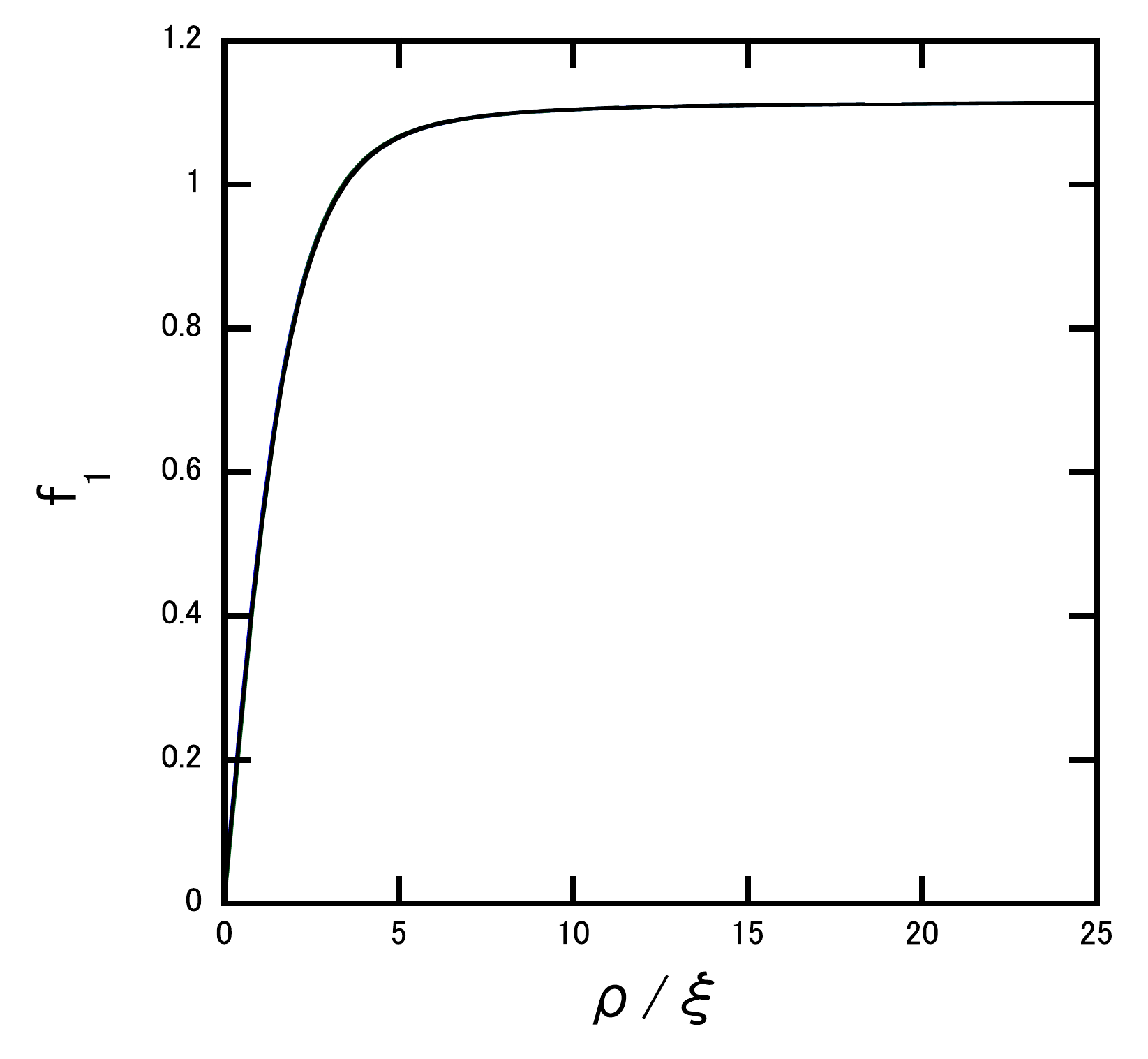}
                  \end{center}
      \end{minipage}
      
      \begin{minipage}{0.33\hsize}
        \begin{center}
          \includegraphics[clip, width=6cm]{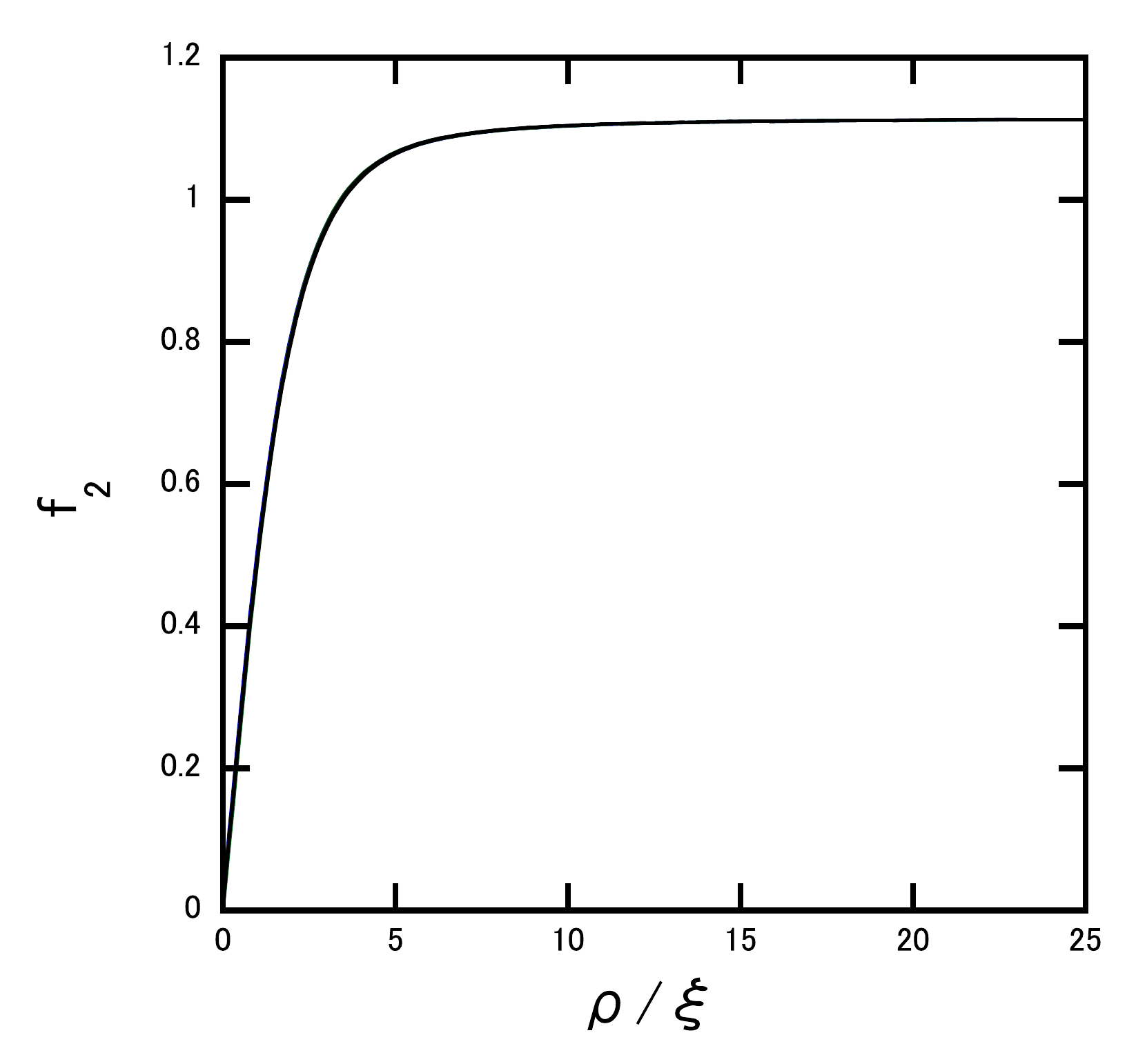}
        \end{center}
      \end{minipage}

      \begin{minipage}{0.33\hsize}
        \begin{center}
          \includegraphics[clip, width=6cm]{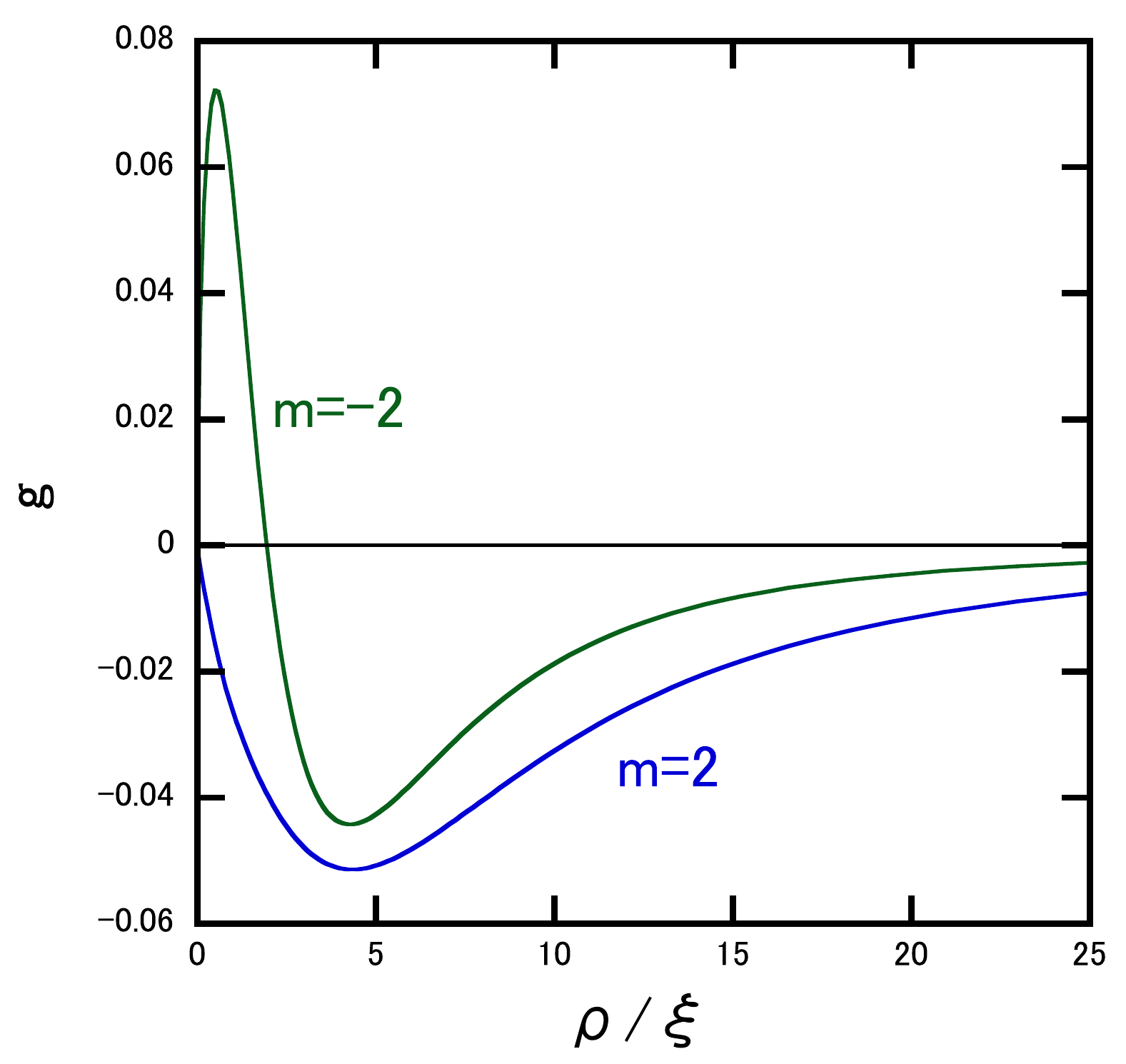}
        \end{center}
      \end{minipage}

    \end{tabular}
    \caption{The profile functions $f_1$, $f_2$ and $g$ as functions of the distance $\rho/\xi$  from the vortex center 
in the case (2). 
The profile functions $f_1$ and $f_2$ are identical: $f_1=f_2$ in this case.
    The blue, green black curves correspond to the cases of 
$g\neq 0$ with $m=2$, $g\neq 0$ with $m=-2$ and $g=0$, respectively. 
All the cases 
take different values numerically although the profiles of $f_1$ and $f_2$ are almost overlapped.}
    \label{fig:case2}
  \end{center}
\end{figure*}

As we discussed before,
if we add the sixth order term,
the ground state is in the UN phase.
Since $f_1=f_2$ is the boundary between the cases with $n=0$ and $n=1$ 
(see Fig.~\ref{condition-f}),
we cannot determine the basis easily.
However, by considering the rotational energy from the basis,
we take $n=0$ in this paper.
Fig.~\ref{fig:case2} shows $f_1$, $f_2$ and $g$ as functions of $\rho$.
In this situation, we find 
from Eq. (\ref{eom-xyz}) in Appendix \ref{sec:eom} that 
$g = 0$ is a trivial solution in the entire region of $\rho$ 
except for the cases of $m  = \pm 2$. 
This implies that all $m (\neq \pm 2)$ give the same solution. 
In the $n=0$ basis,
the equation of motions for $f_1$ and $f_2$ have the same form.
Since $f_1=f_2$ is satisfied at the boundary as we have already mentioned,
$f_1=f_2$ holds at all $\rho$ for any $m$.
The blue and green curves correspond to 
the cases of $g\neq0$ with $m=2$ and $m=-2$, respectively,
 while the black curves represent for the case of $g=0$.
In the figures for $f_1$ and $f_2$,
we cannot see the differences among all the cases 
but they have tiny differences numerically. 

\begin{figure*}[htbp]
  \begin{center}
    \begin{tabular}{c}

      \begin{minipage}{0.33\hsize}
        \begin{center}
          \includegraphics[clip, width=6cm]{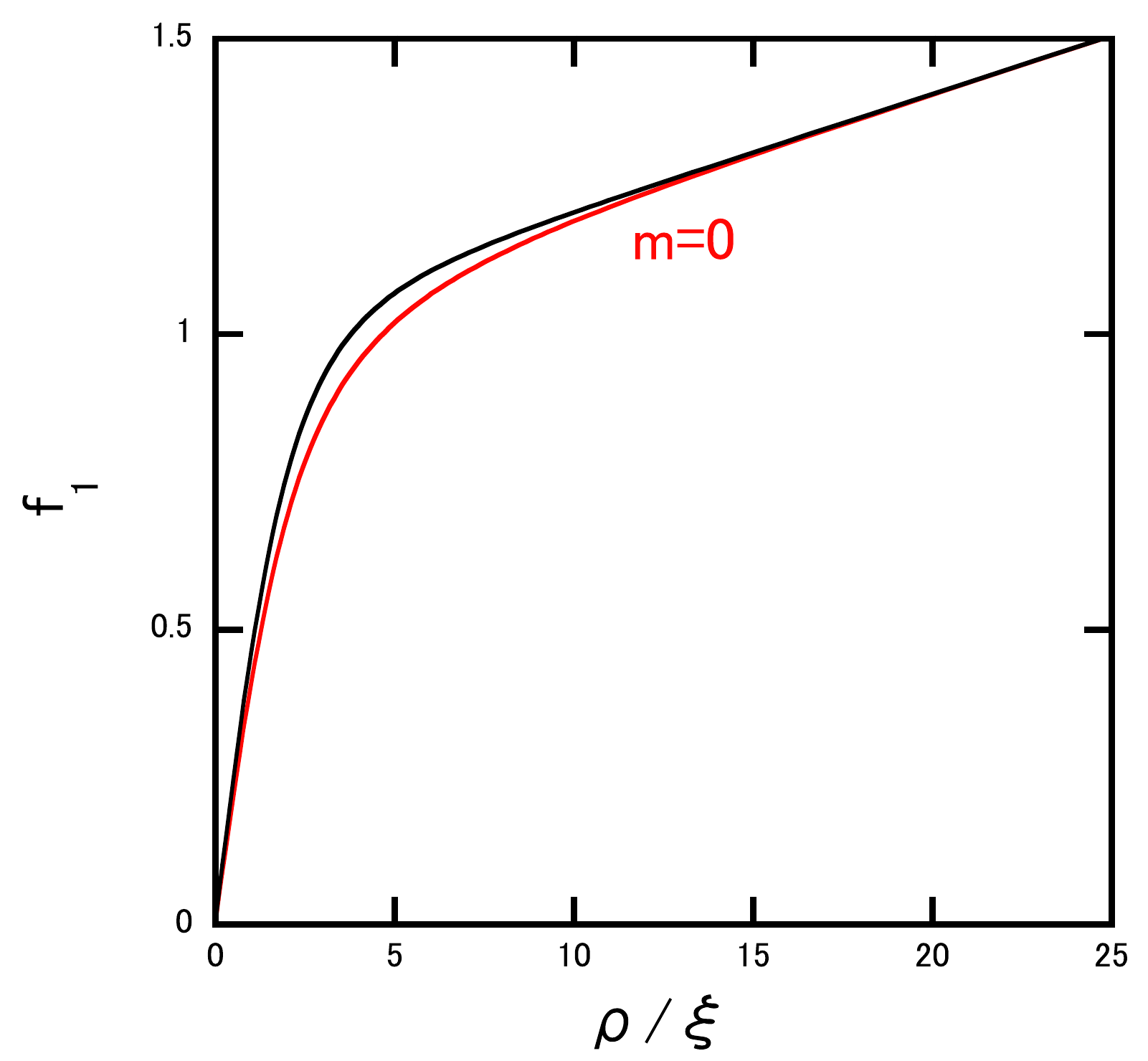}
                  \end{center}
      \end{minipage}
      
      \begin{minipage}{0.33\hsize}
        \begin{center}
          \includegraphics[clip, width=6cm]{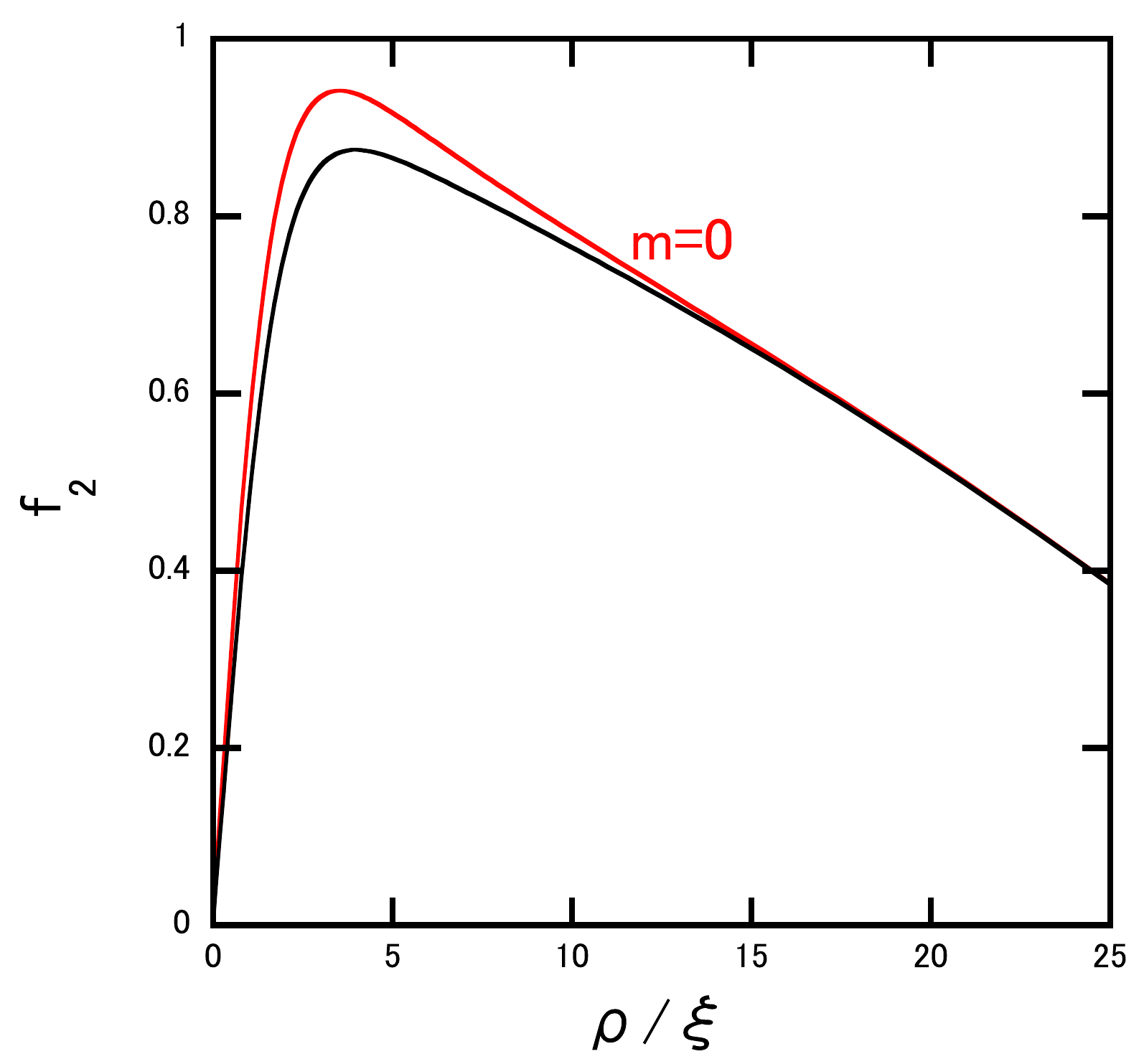}
        \end{center}
      \end{minipage}

      \begin{minipage}{0.33\hsize}
        \begin{center}
          \includegraphics[clip, width=6cm]{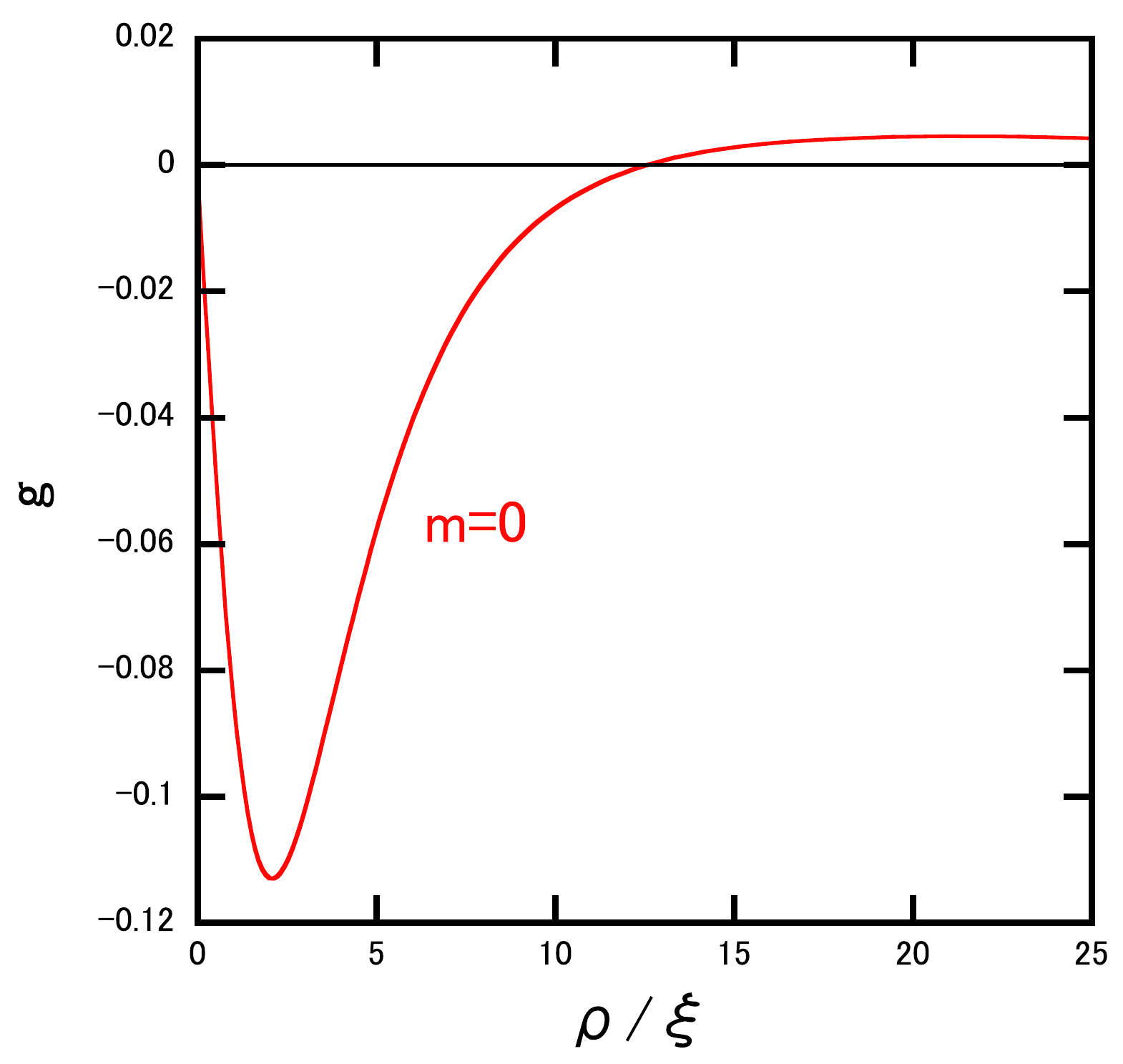}
        \end{center}
      \end{minipage}

    \end{tabular}
    \caption{The profile functions $f_1$, $f_2$ and $g$ as functions of the distance $\rho/\xi$ from the vortex center in the case (3a).
    The red and black curves correspond to the cases of $g \neq 0$ with $m=0$ and $g=0$, respectively.}
    \label{fig:case3}
  \end{center}
\end{figure*}

\begin{figure*}[htbp]
  \begin{center}
    \begin{tabular}{c}

      \begin{minipage}{0.33\hsize}
        \begin{center}
          \includegraphics[clip, width=6cm]{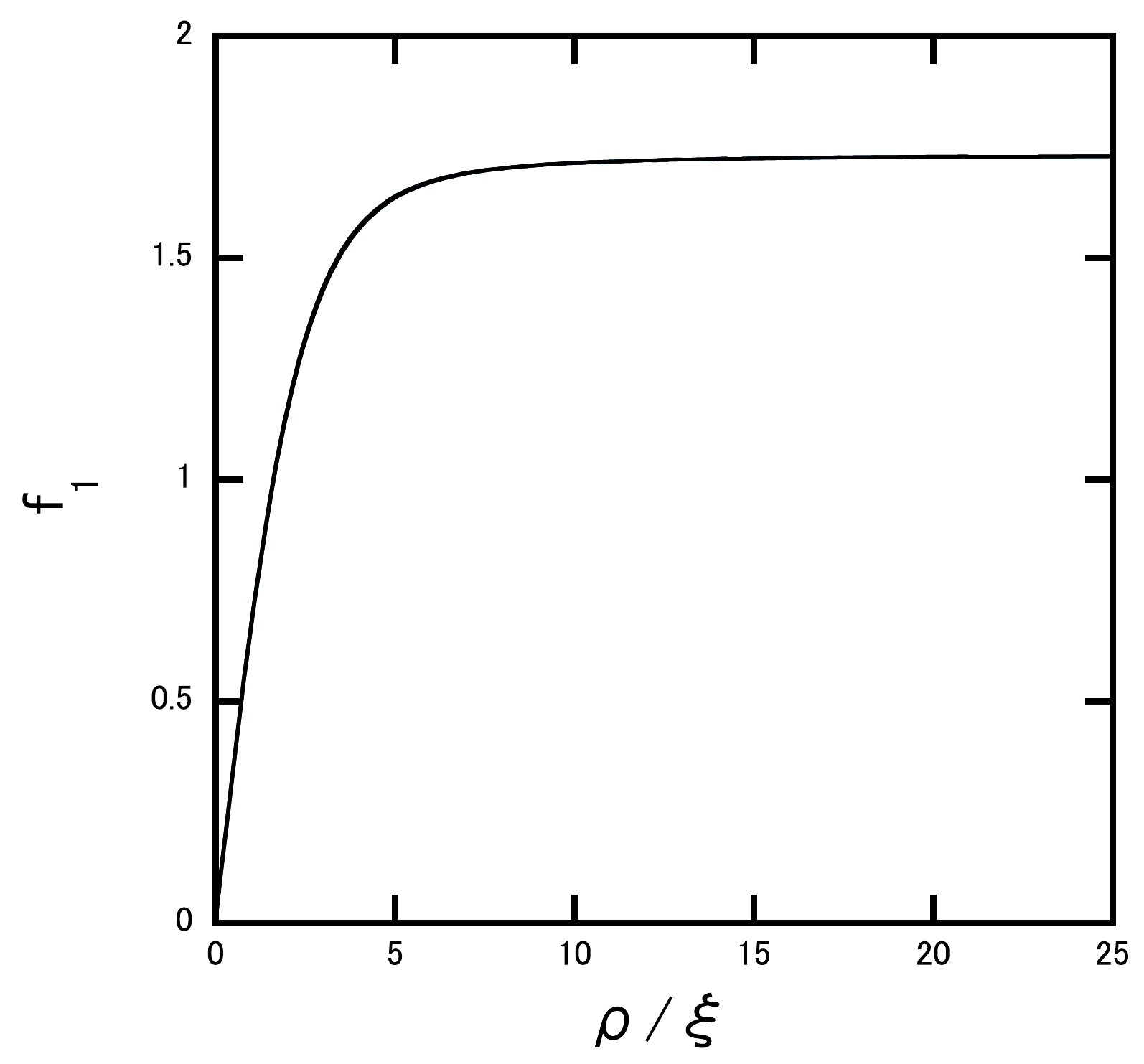}
                  \end{center}
      \end{minipage}
      
      \begin{minipage}{0.33\hsize}
        \begin{center}
          \includegraphics[clip, width=6cm]{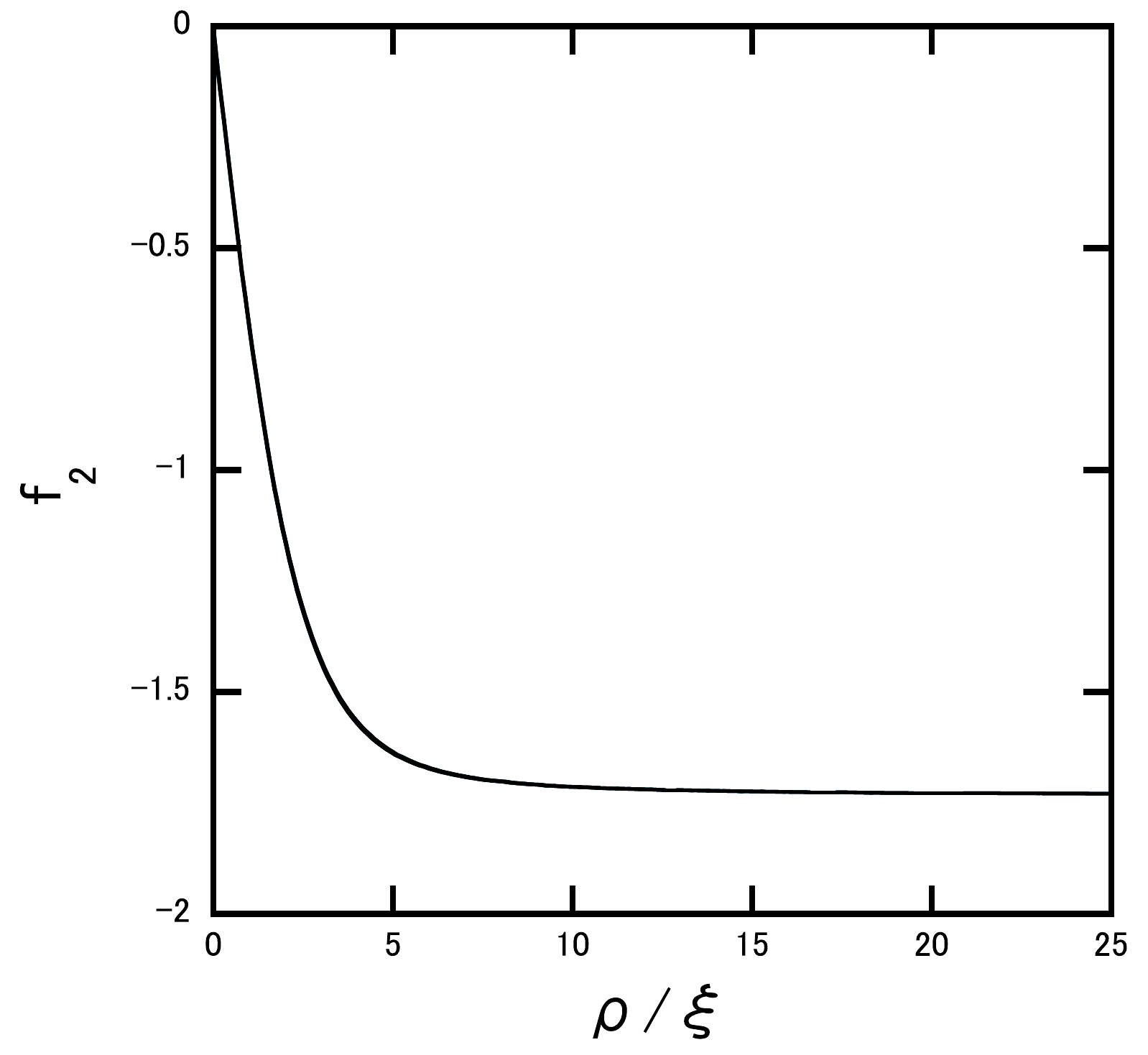}
        \end{center}
      \end{minipage}

      \begin{minipage}{0.33\hsize}
        \begin{center}
          \includegraphics[clip, width=6cm]{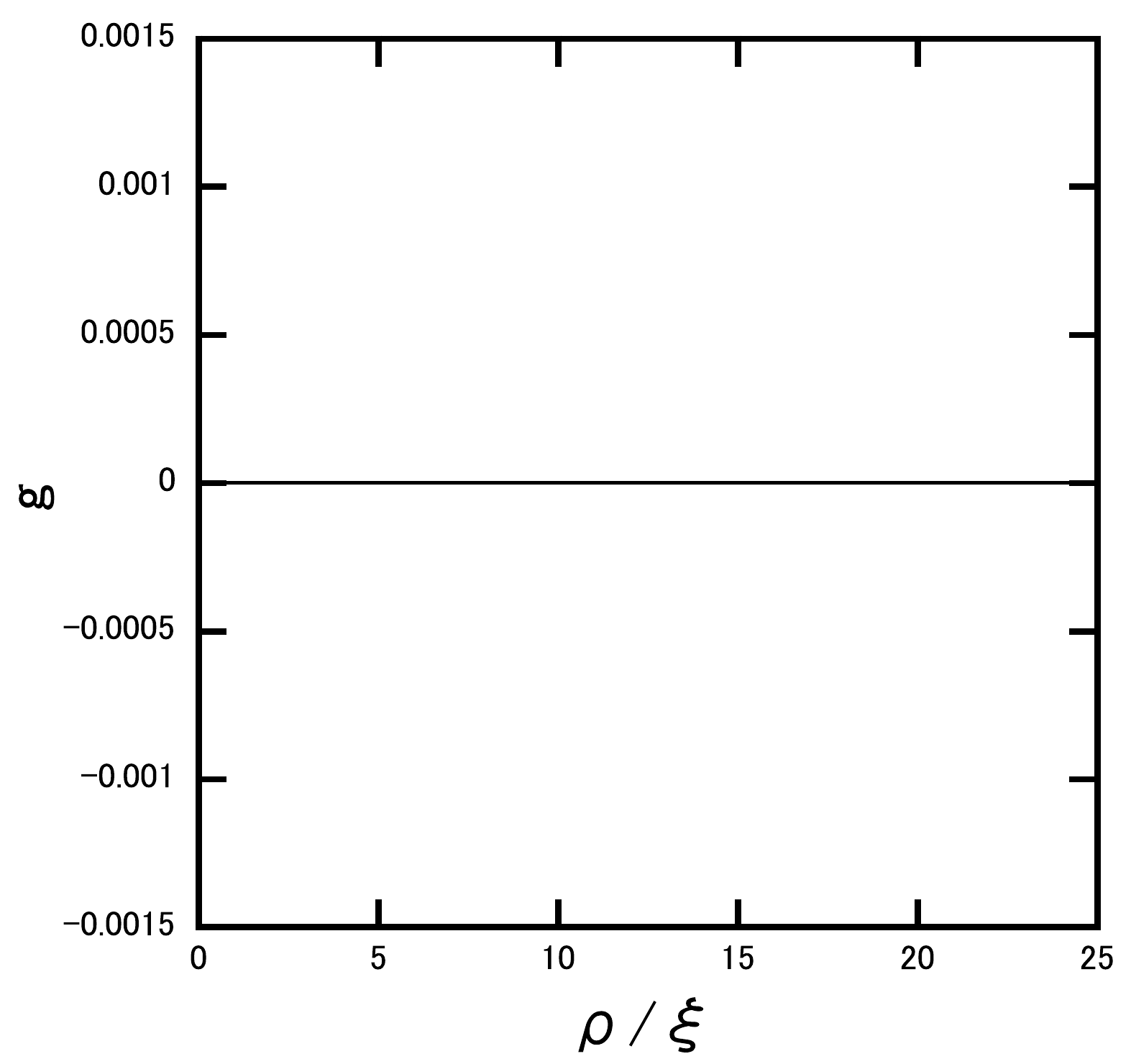}
        \end{center}
      \end{minipage}

    \end{tabular}
    \caption{The profile functions $f_1$, $f_2$ and $g$ as functions of the distance $\rho/\xi$ from the vortex center in the case (3b).
    All $m$ result in the same solution.
}
    \label{fig:case4}
  \end{center}
\end{figure*}

\begin{figure*}[htbp]
  \begin{center}
    \begin{tabular}{c}

      \begin{minipage}{0.33\hsize}
        \begin{center}
          \includegraphics[clip, width=6cm]{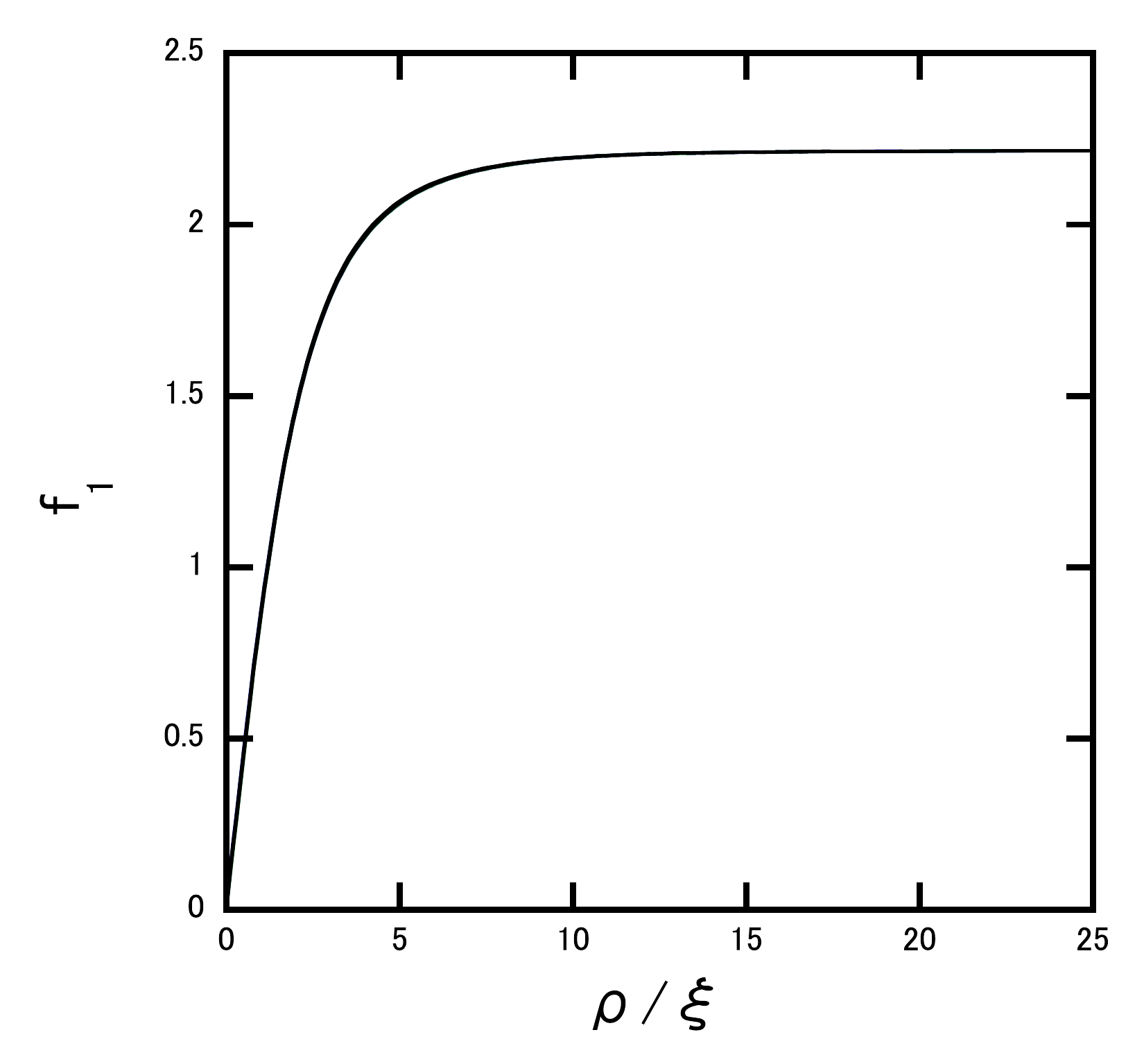}
                  \end{center}
      \end{minipage}
      
      \begin{minipage}{0.33\hsize}
        \begin{center}
          \includegraphics[clip, width=6cm]{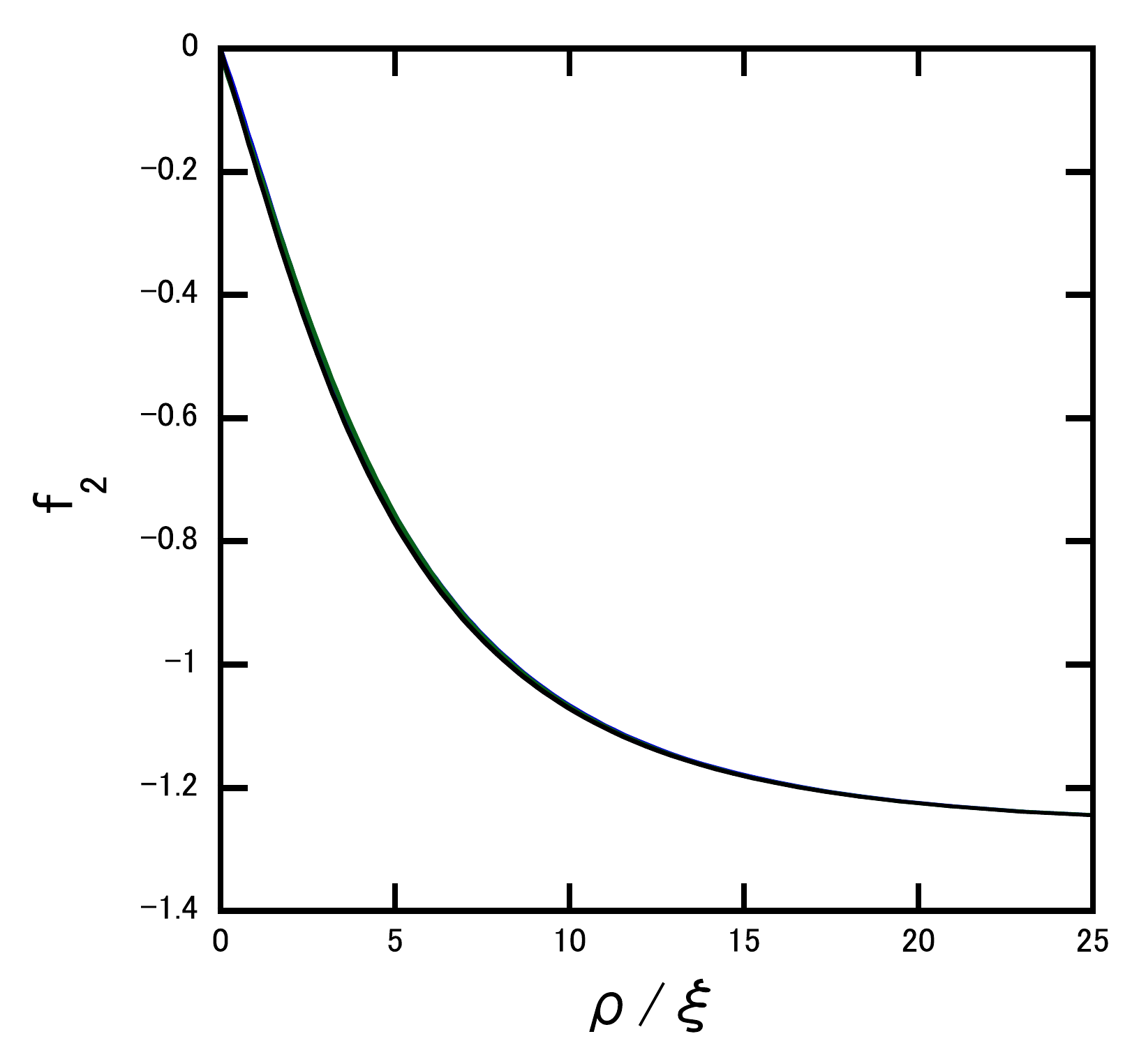}
        \end{center}
      \end{minipage}

      \begin{minipage}{0.33\hsize}
        \begin{center}
          \includegraphics[clip, width=6cm]{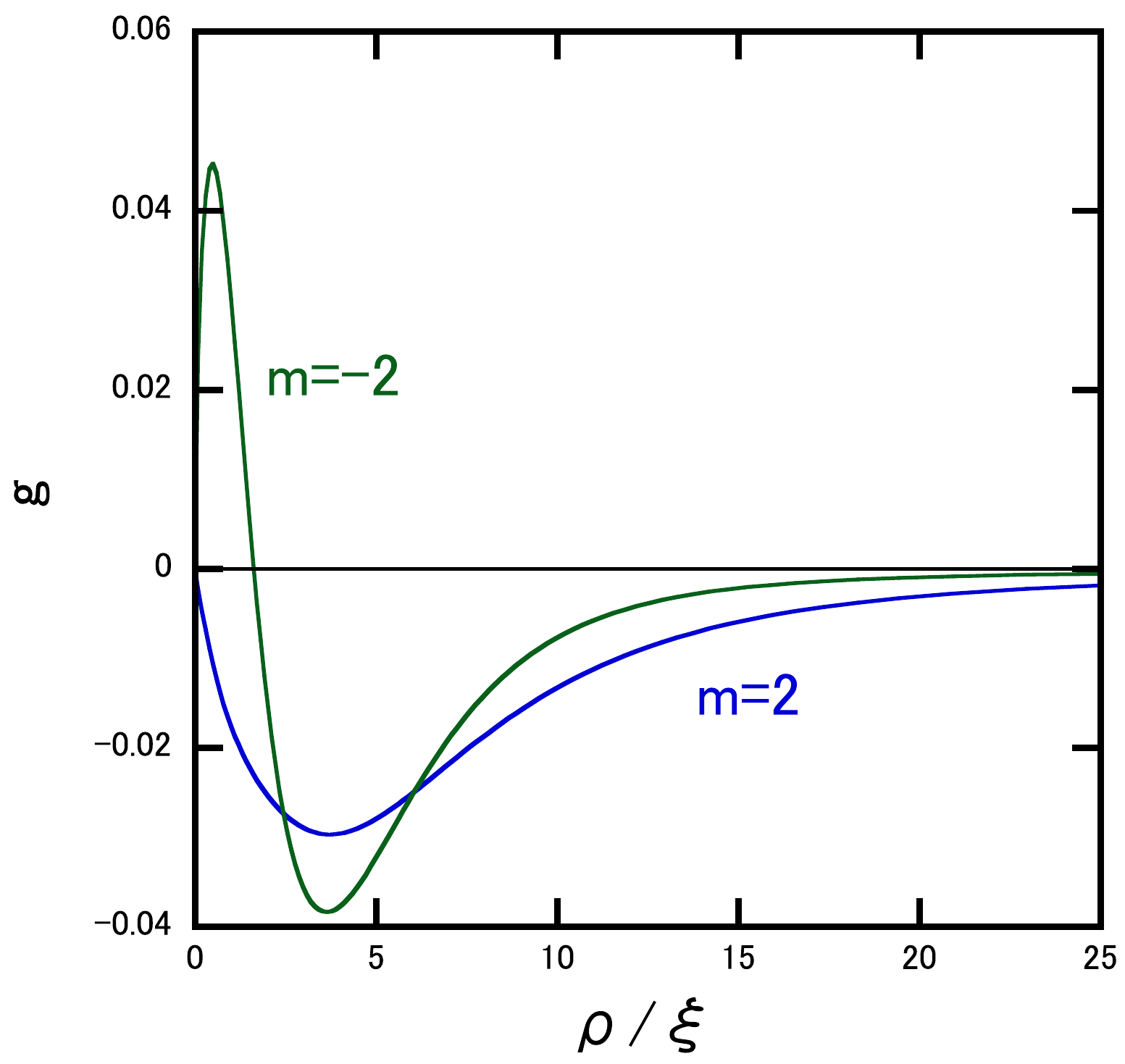}
        \end{center}
      \end{minipage}

    \end{tabular}
    \caption{The profile functions $f_1$, $f_2$ and $g$ as functions of the distance $\rho/\xi$ from the vortex center in the case (4).
   The blue, green and black curves correspond to the cases of 
$g\neq 0$ with $m=2$, $g\neq 0$ with $m=-2$ 
and $g=0$, 
respectively.
    In the figures of $f_1$ and $f_2$,
   all the cases take the different values numerically.}
    \label{fig:case5}
  \end{center}
\end{figure*}

\bigskip
\underline{Cases (3a) and (3b)}

Let us consider the cases that external magnetic fields are present 
either along the $\bm{\theta}$ direction encircling the vortex [the case (3a)]  
or  along the vortex direction [the case (3b)]. 

When the magnetic field along the $\bm{\theta}$ direction is present,
the $D_4$ BN phase in the cylindrical ($n=1$) basis is realized as 
the boundary state. 
We plot $f_1$, $f_2$ and $g$ as functions of $\rho$ in 
Fig.~\ref{fig:case3}.
In the $n=1$ basis,
from the same reason as the case (1),
only the case with $m=0$ has a non-zero value for $g$.
The red and black curves correspond to the cases of 
 $g \neq 0$ (with $m=0$) and $g=0$, respectively.

When the magnetic field along the $\bm{z}$ axis is present,
we obtain the $D_4$ BN phase as the boundary state of the vortex as shown 
in Eq.~(\ref{magnetic field}).
Since $f_1=-f_2$ holds at the boundary $\rho \to \infty$, 
we find  from Fig.~\ref{condition-f} 
that the boundary state is in the Cartesian ($n=0$) basis.
Fig.~\ref{fig:case4} shows the profile functions 
$f_1$, $f_2$ and $g$ as functions of $\rho$ 
in the case (3b).
Since the relation $f_1=-f_2$ is satisfied at the two boundaries and the equation of motions for $f_1$ and $f_2$  take the same form, 
the relation $f_1=-f_2$ holds in the entire region of $\rho$.
Consequently,
the terms proportional to $\frac{\partial^2 f_{1(2)}}{\partial \rho^2}$ and $\frac{\partial f_{1(2)}}{\rho \partial \rho}$ vanish.
As a result, $g$ has a trivial solution $g=0$ for all $m$,
implying that all $m$ give the identical solution. 

In the cases (3a) and (3b), the ground state does not depend on the magnitude of the magnetic field,
while the profile of vortices of course depends on the magnitude of the magnetic field.
We set the magnetic field to be 
$10^{15}$ Gauss, which corresponds to magnetars.
To realize the situations in the cases (3a) and (3b),
we need a strong magnetic field larger than about $3\times 10^{15}$ Gauss 
as can be seen from Fig.~\ref{condition-f}.
Therefore, the value $10^{15}$ Gauss is not appropriate, 
but the qualitative behaviors of vortex profiles 
are not changed even 
when we change the magnitude
of the magnetic field to  $10^{16-17}$ Gauss.

\bigskip
\underline{Case (4)}

Finally, let us consider 
the most realistic case for neutron stars, that is, 
the case with the sixth order term in the presence of the magnetic field 
of $10^{15}$ Gauss along the $\bm{z}$ axis.
This case reduces to the case (2) in the absence of 
the magnetic field and to the case (3b) in the presence of strong magnetic field 
for which the sixth order term is negligible.

In this case, the smallest eigenvalue $-1-r$ comes to the $z$ component.
By using Fig.~\ref{condition-f},
we can see that the  $xyz$ basis ($n=0$) is realized.
By minimizing Eq.~(\ref{6th+mag}),
we have $r = r_{\rm tot} \sim -0.572$ in Eq.~(\ref{eq:rtot}) 
as the boundary condition at large distance.

In Fig.~\ref{fig:case5}, 
we plot $f_1,\ f_2$ and $g$ as functions of $\rho$ with $m=-2,...,2$ 
in the case (4).
For the same reason as the case (2),
only the cases with $m = \pm 2$ have 
a non-zero value for $g$.
The blue, green and black curves correspond to the cases of 
$g \neq 0$  with $m=2$, $g \neq 0$ with $m=-2$,  
and $g =0$, respectively.
In figures for $f_1$ and $f_2$,
the cases of $g \neq 0$ with $m= \pm 2$ 
and of $g=0$ take the different values numerically 
although they are almost overlapped.

\section{Spontaneous magnetization of the $^3P_2$ vortex core}\label{sec:magnetization}

\begin{figure*}[htbp]
  \begin{center}
    \begin{tabular}{c}

      \begin{minipage}{0.33\hsize}
        \begin{center}
          \includegraphics[clip, width=6cm]{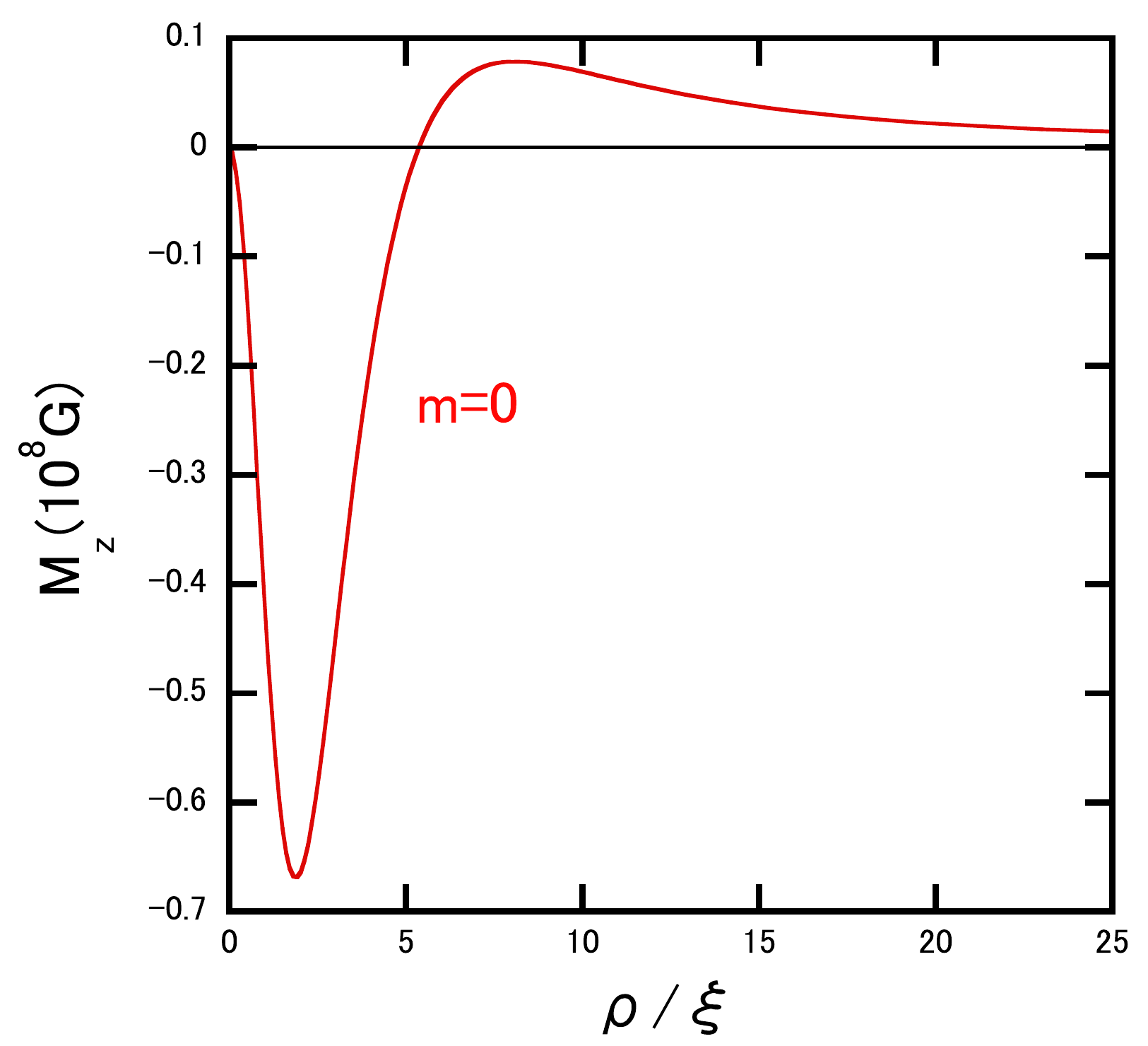}
                  \end{center}
      \end{minipage}
      
      \begin{minipage}{0.33\hsize}
        \begin{center}
          \includegraphics[clip, width=6cm]{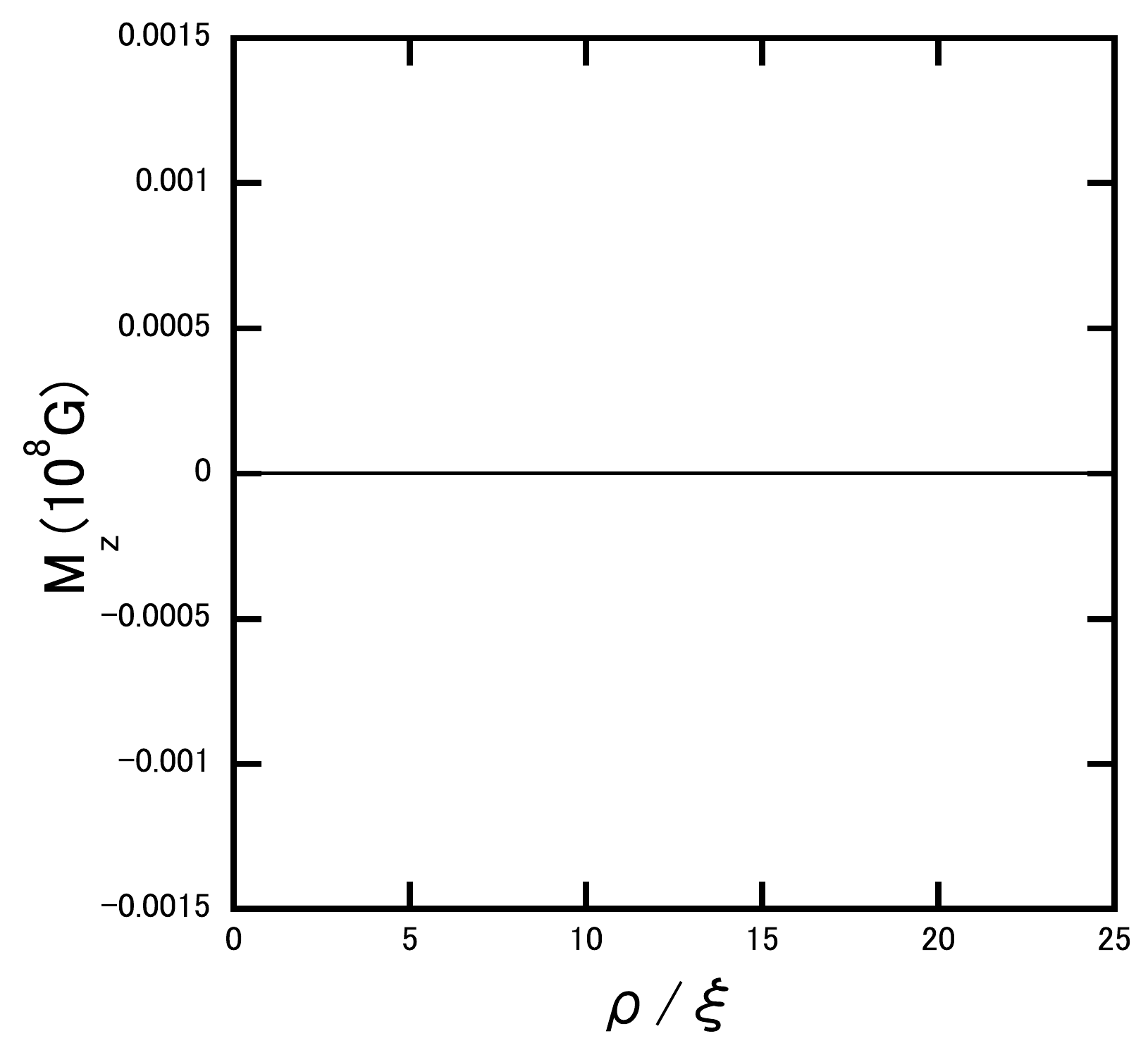}
        \end{center}
      \end{minipage}

      \begin{minipage}{0.33\hsize}
        \begin{center}
          \includegraphics[clip, width=6cm]{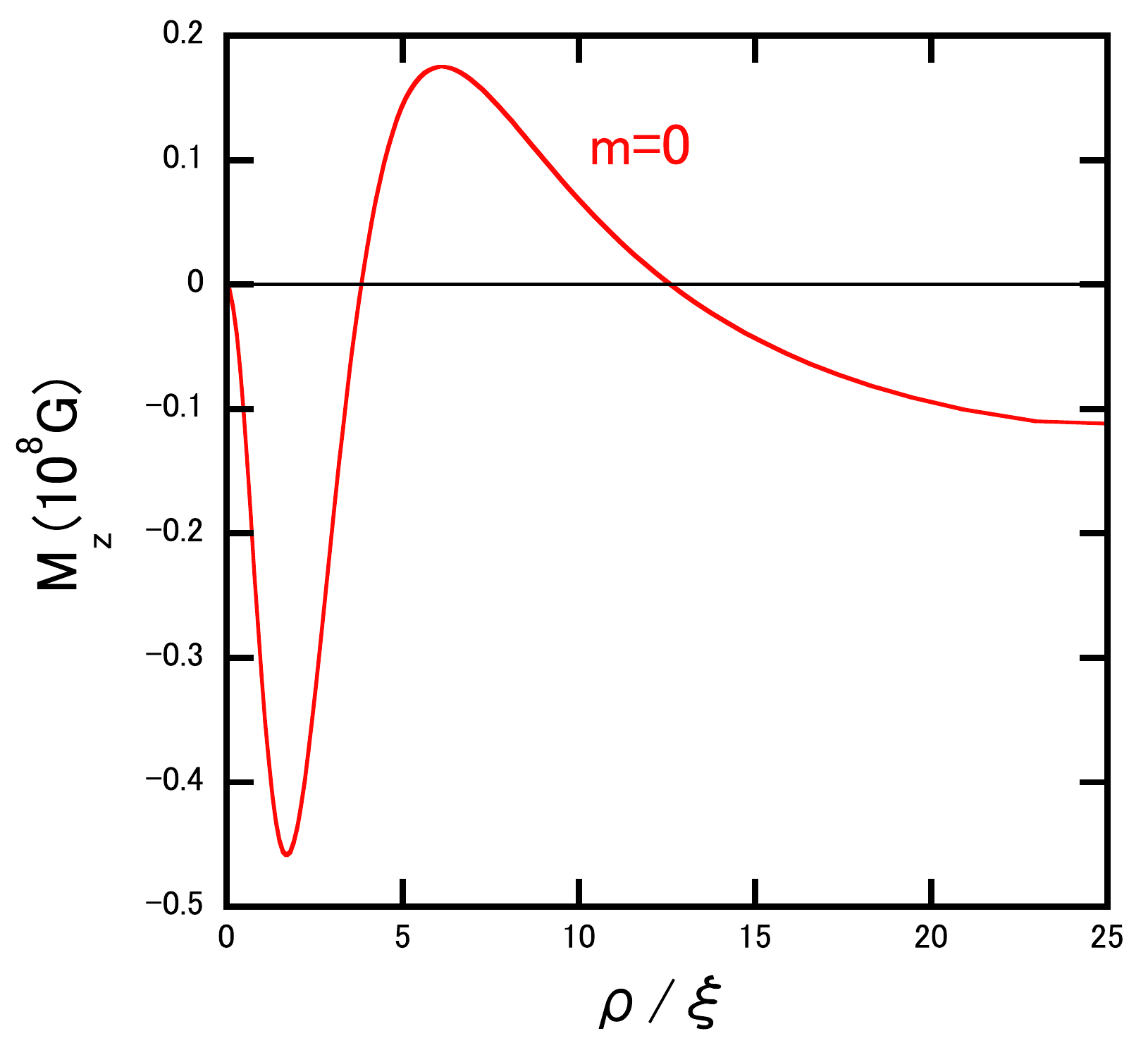}
        \end{center}
      \end{minipage}
\\

      \begin{minipage}{0.33\hsize}
        \begin{center}
          \includegraphics[clip, width=6cm]{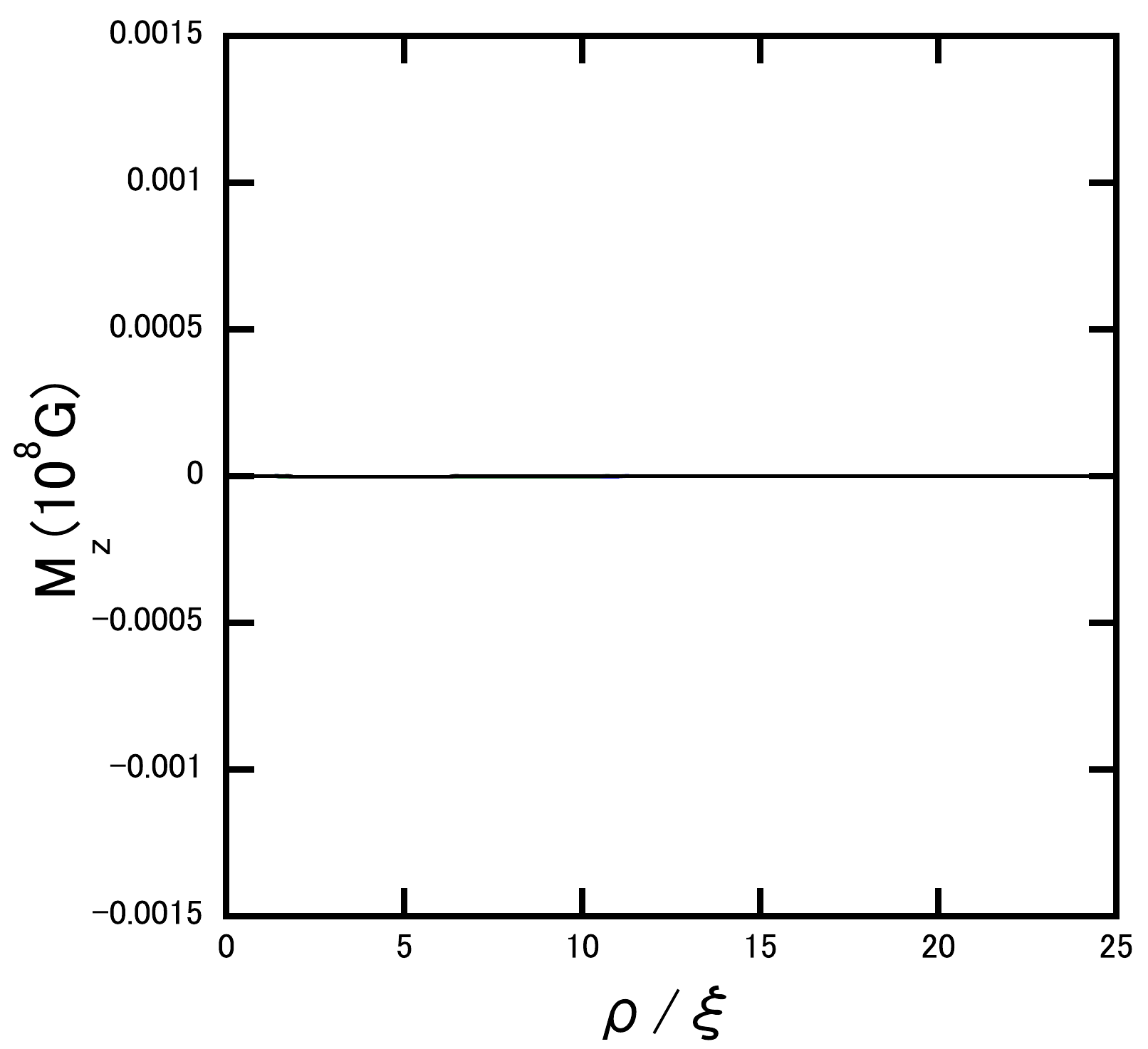}
                  \end{center}
      \end{minipage}
      
      \begin{minipage}{0.33\hsize}
        \begin{center}
          \includegraphics[clip, width=6cm]{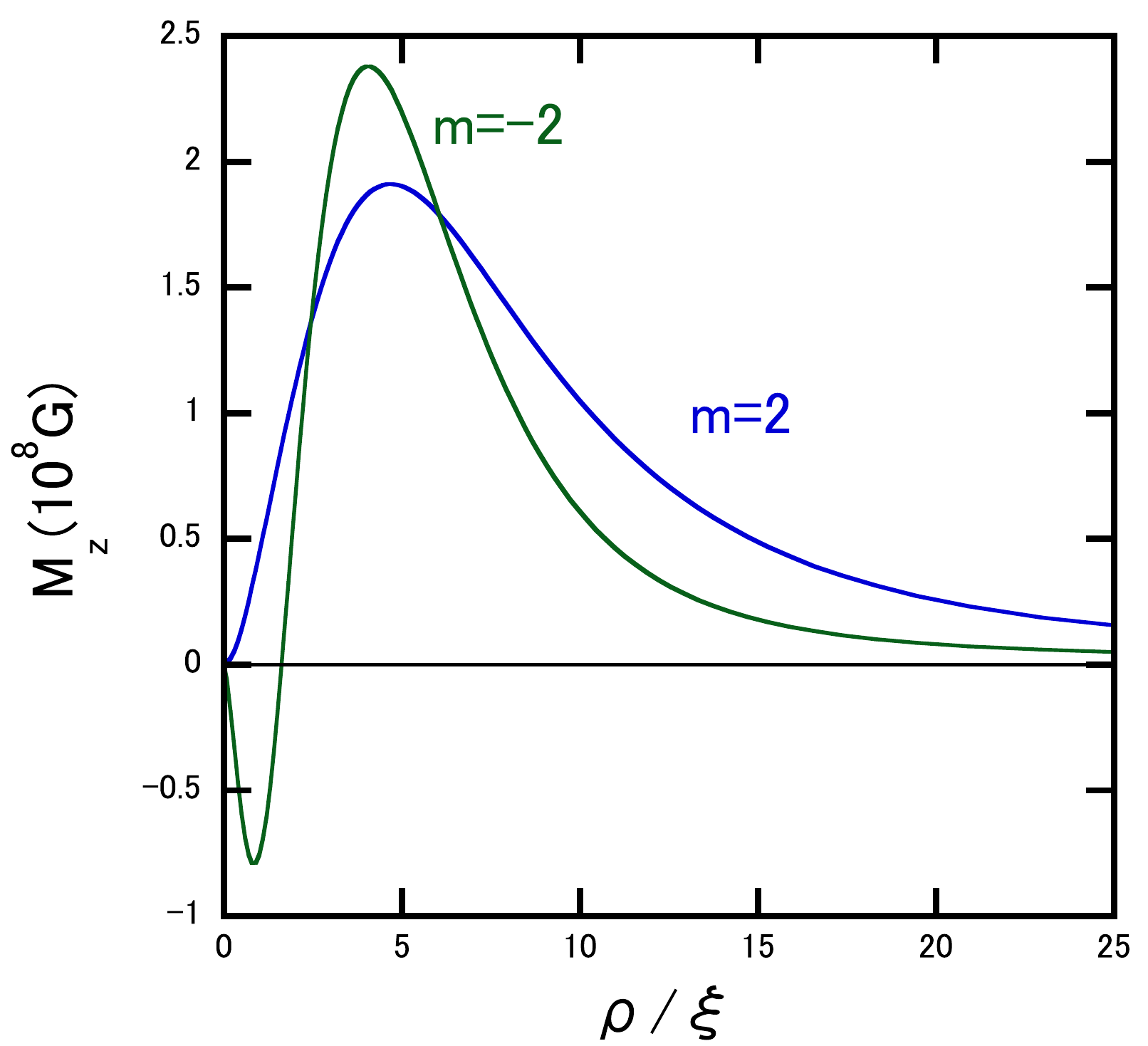}
        \end{center}
      \end{minipage}
    \end{tabular}
    \caption{The dependence of the magnetizations $M_z$ 
on the distance $\rho/\xi$ from the vortex core 
in the cases (1), (2), (3a), (3b) and (4).
    The red, blue and green curves correspond to the cases of 
$m=0$, $m=2$ and $m=-2$ with $g\neq 0$, respectively, 
while the black lines correspond to the case with $g=0$ in the 
cases (1), (3a), (3b) and (4). 
The magnetization $M_z$ vanishes in the case (2) 
because of $f_1=f_2$ even for $g \neq 0$.
} 
    \label{fig:M}
  \end{center}
\end{figure*}
In this section, 
we calculate the spontaneous magnetization of $^3P_2$ vortex cores 
due to the neutron anomalous magnetic moment 
for
the vortex profiles obtained in the last section. 
The spontaneous magnetization was already reported in 
the case (1) \cite{Sauls:1982ie}. 
This is a characteristic feature of $^3P_2$ vortices 
that is absent for conventional $^1S_0$ vortices.
The vortex magnetization $\bm{M}(\rho)$ can be calculated as 
\begin{eqnarray}
\bm{M}&=&\frac{\gamma_n \hbar}{2} \bm{\hat{ \sigma}}, \nonumber \\
\bm{ \hat{\sigma}}&=& T\sum_n \int \frac{d^3k}{(2\pi)^3}{\rm Tr}(\bm{\sigma}G(k,\omega_n))  \nonumber \\
&=&\int \frac{d\Omega}{4\pi} {\rm Tr}(\bm{\sigma}\Delta \Delta^{\dagger})T\sum_n\int d\xi N(0) \frac{i\omega_n+\xi}{(\omega_n^2+\xi^2)^2}  \nonumber \\
&=&\frac{4}{9}N'(0)k_F^2\frac{|\alpha|}{6\beta}g(\rho)(f_1(\rho)-f_2(\rho)){\rm cos}m\theta \hat{\bm{z}}
\end{eqnarray}
where $G(k,\omega_n)$ is a thermal Green function and $\omega_{n}=(2n +1)\pi T$ is the Matsubara frequency 
and $N'(0) = {M^2 \over 2\pi^2 k_F}$ is the density of states differentiated 
by the energy $E=k^2/2M$, $N' = {M^2 \over 2\pi^2 k}$, 
evaluated at the Fermi surface $k=k_F$.

By using the results of the last section, 
we obtain the magnetization $\bm{M}$ as a function of $\rho$.
We plot $M_z$ (for $\theta=0$) as 
functions of $\rho$ for the cases (1)--(4) in Fig.~\ref{fig:M}. 
The red, blue and green curves correspond to the cases of $g \neq 0$ 
with $m=0$, $m=2$ and $m=-2$, respectively, 
while the black curves correspond to the case of $g=0$.
The maximum value of $M_z$ is about $10^8$-$10^9$ Gauss 
when it is nonzero. 
The mean magnetic field in a vortex lattice,  
which is  
obtained roughly by multiplying $(\xi/d)^2 \sim 10^{-14}$, 
is much smaller than the observed magnetic field 
about $10^{12}$-$10^{15}$ Gauss,
and consequently this magnetic field is negligible.

Note that the magnetization $M_z$ 
is proportional to $(f_1(\rho) - f_2(\rho)) g (\rho)\,{\rm cos}m\theta$.
Since it is proportional to 
the off-diagonal profile function $g$ 
appearing around the vortex cores,  
it is nonzero only for the cases of $m=0$ in the $xyz$-basis
and of  $m = \pm 2$ in the cylindrical basis.
When $f_1 \neq f_2$ in the region $g \neq 0$, 
the magnetization can occur.
However, 
in the case (2), as we have already discussed,
$f_1=f_2$ is satisfied for all $\rho$,  
and the magnetization $M_z$ vanishes.  
Among all the cases with nonzero magnetization $M_z$, 
only the case with $m=0$ has a net magnetization, 
since  the $\theta$ integration of ${\rm cos}m\theta$ vanishes for $m \neq 0$. 
Although the case of   $m = \pm 2$ in the cylindrical basis 
has no net magnetization, 
the direction of the magnetization drastically changes 
upward and downward depending on $\theta$, 
thereby implying the existence of the large current 
crossing to the vortex.

\section{Summary and Discussion} \label{sec:summary}

We have determined the ground states 
of the $^3P_2$ superfluids in the presence of 
the external magnetic fields 
and have obtained the vortex solutions 
in various situations in the absence and presence of magnetic fields 
along the vortex axis or the angular direction. 
First, the boundary state at $\rho \rightarrow \infty$ 
has been determined to lower the bulk free energy.
Second, the basis diagonalizing the order parameter $A_{\mu i}$ 
has been determined by the leading contribution to the gradient energy proportional to 
log $L$ and the coefficients $K_1$ and $K_2$.
The presence of the gradient energy proportional to $K_2$ is essential 
for the $^3P_2$ superfluids unlike the spinor BEC for which 
such the terms are absent.  
In the absence of the magnetic field and the sixth order term, 
the ground state is continuously degenerate with the parameter $r$. 
Nevertheless the boundary value $r$ for the vortex state is fixed 
and the cylindrical basis is realized 
due to the gradient energy.  
We have constructed the free energy density and 
equation of motion for the $xyz$-basis and cylindrical basis.
By using the Ansatz of the vortex profiles 
with the non-diagonal component $g$ 
which has the local winding number $m$ relative to that of $f_{1,2}$,  
we have constructed the vortex profiles and 
have found that $g$ is non-zero around the vortex core  
only for $m=0$ in the cylindrical basis and $m = \pm 2$ in the $xyz$-basis. 
The former was known before without explicit profiles 
\cite{Richardson:1972xn,Muzikar:1980as,Sauls:1982ie} for which 
we have given the explicit solution.  
As a result,  the spontaneous magnetization around the vortex cores 
proportional to $g\, {\rm cos}m\theta$ is present 
for these cases. 
The net magnetization survives for $m=0$ in the cylindrical basis,
while in the case of  $m = \pm 2$ in the $xyz$-basis 
the net magnetization vanishes with local magnetization upward and downward 
depending on the angle coordinates. 
The typical value for the spontaneous magnetization is about $10^{8}$ Gauss,
which is much smaller that the neutron star observations.

Here we address some discussion.
The unbroken symmetry is $D_4$ for 
the $D_4$ BN phase realized in the cases (3a) and (3b).
Since the element of the $D_4$ group consists of an element of simultaneous 
action of a  $U(1)$ phase rotation by $\pi$ and an $SO(3)$  rotation  by $\pi/2$,
the most fundamental vortex is a half-quantized vortex.
In the BN phase,  
a single integer vortex discussed in this paper may be 
unstable against the decay into two half quantized vortices.
The half quantized vortices belong to 
the non-Abelian first homotopy group $D_4^*$ as in Table \ref{table-sym}.
When two vortices characterized by elements that do not commute with each other
collide, a bridge between them must be created \cite{Mermin:1979zz},
see Ref.~\cite{Kobayashi:2008pk} for such a collision dynamics in 
spinor BEC. 
We should consider the effects of the half quantized vortices 
in the $^3P_2$ neutron superfluids. 

We have discussed only vortices characterized by the first homotopy group.
On the other hand, 
in Table \ref{table-sym}, we summarized higher homotopy groups 
that allow higher codimensional topological objects like 
monopoles, Skyrmions and so on. 

It was argued in \cite{Nitta:2013mj} that 
in helium 3 superfluid, 
the $SO(3)$ symmetry unbroken in the ground state 
is further spontaneously broken around the core of 
an integer vortex due to the gradient term proportional to $K_2$, 
giving rise to (quasi-) gapless mode localized along the vortex.
The same discussion may be made for the integer vortex 
in the $^3P_2$ superfluids studied in this paper.

In order to study the fermionic degree of freedom, 
the Bogoliubov-de-Gennes equation should be used 
beyond the GL free energy.
It will turn out to be useful to show 
topological properties of
the $^3P_2$ superfluidity such as topological superfluidity 
and fermion zero modes trapped inside the vortex core. 

The interface between $^1 S_0$ and $^3P_2$ superfluids 
should be important to consider; in particular how vortices are connected 
in the both region. On the other hand, at much higher density, 
quark matter may appear such as color superconductors \cite{Alford:2007xm}, 
where there exist 1/3 quantized superfluid vortices that carry 
color magnetic fluxes  
\cite{Balachandran:2005ev,Nakano:2007dr,Eto:2009kg,Eto:2009tr,Eto:2013hoa},
the boojum structure on which vortices join will 
appear at the interface \cite{Cipriani:2012hr}. 
Dynamics of vortices at such the interfaces ($^1 S_0$-$^3P_2$ 
and $^3P_2$-quark matter) may be important for 
dynamics of neutron stars. 

\begin{acknowledgments}

We thank Tatsuyuki Takatsuka, Tetsuo Hatsuda, Takeshi Mizushima and
James Avery Sauls for helpful discussions and comments.
KM thanks Mark Alford for the kind hospitality and discussions in Washington University in St.~Louis
where part of this work was carried out under the support of ALPS Program, 
University of Tokyo.
KM is supported by JSPS Research Fellowship for Young Scientists.
The work of M.~N.~is supported in part by a Grant-in-Aid for
Scientific Research on Innovative Areas ``Topological Materials
Science'' (KAKENHI Grant No.~15H05855) and ``Nuclear Matter in Neutron
Stars Investigated by Experiments and Astronomical Observations''
(KAKENHI Grant No.~15H00841) from the the Ministry of Education,
Culture, Sports, Science (MEXT) of Japan. The work of M.~N.~is also
supported in part by the Japan Society for the Promotion of Science
(JSPS) Grant-in-Aid for Scientific Research (KAKENHI Grant
No.~25400268) and by the MEXT-Supported Program for the Strategic
Research Foundation at Private Universities ``Topological Science''
(Grant No.~S1511006).

\end{acknowledgments}

\appendix

\section{Ground state with the sixth order term} \label{sec:ground_state}

In this appendix, we determine the ground state taking into account 
the sixth oder term in the GL free energy. 
We correct the amplitude 
of the state with the sixth order term previously obtained in Ref.~\cite{Sauls:1978lna}, 
but the difference is negligible.  
We also show that off-diagonal elements do not appear, and  
so the state remains in the nematic phase even when we take into account the 
sixth order term.

We consider the following free energy;
\begin{eqnarray}
F=\int d^3 \rho \ (f_4+f_6 ).
\end{eqnarray}
By minimizing $f_4$,
we can obtain the ground state up to fourth order $A_{{\rm 4th}}$,
\begin{eqnarray}
A_{{\rm 4th}}=\sqrt{\frac{|\alpha|}{6\beta}} 
  \left(
    \begin{array}{ccc}
    1 & 0  & 0 \\
    0  &1 & 0  \\
    0  & 0 & -2 
    \end{array}
  \right).
\end{eqnarray} 
To see the ground state up to the sixth order, we expand the order parameter $A$ as follows:
\begin{eqnarray}
A&=&A_{{\rm 4th}}+ \Delta A \nonumber \\
&=&A_{{\rm 4th}}+i\sqrt{\frac{|\alpha|}{6\beta}} 
  \left(
    \begin{array}{ccc}
    f_1 & g  & h \\
    g  &f_2 & j  \\
    h  & j & -f_1-f_2 
    \end{array}
  \right).
\end{eqnarray}
Then, we put this order parameter into the free energy density and 
expand it up to the second order 
with respect to the $\Delta A$:
\begin{eqnarray}
F=&&\int d^3 \rho \ \bigg[ 
\frac{\alpha^2}{18\beta}\left((f_1-f_2)^2+4g^2+10h^2+10j^2\right) \nonumber \\
&&\quad\quad\quad +\gamma\frac{|\alpha|^3}{216\beta^3}(1704f_1^2+1704f_2^2+768f_1f_2 \nonumber \\
&&\quad \quad \quad+2640g^2+5808h^2+5808j^2) 
\bigg].
\end{eqnarray}
Therefore, if we take $f_1=f_2$, we can obtain the lower free energy density 
since $\gamma$ is negative.
Next, we assume that the ground state up to the sixth order to be 
\begin{eqnarray}
A_{6{\rm th}} = N_{{\rm 6th}}
  \left(
    \begin{array}{ccc}
      1 & 0 & 0 \\
      0 & 1 & 0 \\
      0 & 0 & -2
    \end{array}
     \right).
\end{eqnarray} 
By minimizing the free energy with respect to $N_{{\rm 6th}}$, we get
\begin{eqnarray}
N_{{\rm 6th}}\equiv \sqrt{\frac{6\beta-\sqrt{(6\beta)^2-2784\alpha \gamma}}{1392|\gamma|}}.
\end{eqnarray}
Finally, let us expand the free energy density by using
\begin{eqnarray}
A&=&A_{{\rm 6th}}+ \Delta' A \nonumber \\
&=&A_{{\rm 6th}}+iN_{{\rm 6th}}
  \left(
    \begin{array}{ccc}
    f_1 & g  & h \\
    g  &f_2 & j  \\
    h  & j & -f_1-f_2 
    \end{array}
  \right).
\end{eqnarray}
We thus reach at 
\begin{eqnarray}
F=\int d^3 \rho \  (c_1(f_1-f_2)^2+c_2g^2+c_3h^2+c_4j^2)
\end{eqnarray}
where each coefficients $c_1$,  $c_2$, $c_3$ and $c_4$ are positive.
Therefore, we have shown that $A_{{\rm 6th}}$ is at least at the local minimum.

\section{Equation of motion} \label{sec:eom} 

\begin{widetext}

In this appendix, we write down the equation of motion
in the cylindrical basis ($n=1$) and  $xyz$-basis ($n=0$).
\begin{itemize}
\item  The equation of motions for cylindrical basis ($n=1$) 
are given as follows:
\end{itemize}

\begin{eqnarray}
&&7\frac{\partial^2 f_1}{\partial \rho^2}+\frac{1}{\rho}\left(5\frac{\partial f_1}{\partial \rho}-4\frac{\partial f_2}{\partial \rho}-2\delta_{m,0}\frac{\partial g}{\partial \rho}\right) 
+\frac{1}{\rho^2}\left(-15f_1+14f_2+22\delta_{m,0}g \right) +3f_1 \nonumber \\
&&-f_1\left(f_1^2+f_2^2+f_1f_2+\frac{5+2\delta_{m,0}}{3}g^2\right)+f_2\left(\frac{1+\delta_{m,0}}{3}g^2\right) \nonumber \\
&&-\frac{|\alpha|}{36\beta^2}\gamma \Bigl(216f_1^5+408f_1^4f_2+672f_1^3f_2^2+528f_1^2f_2^3+264f_1f_2^4  
+g^2((972+516\delta_{m,0})f_1^3 \nonumber \\
&&+(504+120\delta_{m,0})f_1^2f_2+(612+300\delta_{m,0})f_1f_2^2-(216+168\delta_{m,0})f_2^3)+g^4( (504+264\delta_{m,0})f_1 \nonumber \\
&&-(144+168\delta_{m,0})f_2)\Bigr) +\frac{g}{|\alpha|} \times \begin{cases}
-(f_1+f_2) H_z^2 \ \ \ \ \   (\bm{H} \parallel \bm{z)} \\
f_2 H_{\theta}^2  \ \ \ \ \   (\bm{H} \parallel \bm{\theta)}
\end{cases}
=0,
\end{eqnarray}
\begin{eqnarray}
&&3\frac{\partial^2 f_2}{\partial \rho^2}-2\frac{\partial^2 f_1}{\partial \rho^2}
+\frac{1}{\rho}\left(2\frac{\partial f_1}{\partial \rho}+5\frac{\partial f_2}{\partial \rho}-2\delta_{m,0}\frac{\partial g}{\partial \rho}\right)
+\frac{1}{\rho^2}\left(12f_1-19f_2-26\delta_{m,0}g\right) +3f_2 \nonumber \\
&&-f_2\left(f_1^2+f_2^2+f_1f_2+\frac{5+2\delta_{m,0}}{3}g^2 \right)+f_1\left(\frac{1+\delta_{m,0}}{3}g^2\right) \nonumber \\
&&-\frac{|\alpha|}{36\beta^2}\gamma \Bigl(216f_2^5+408f_2^4f_1+672f_2^3f_1^2+528f_2^2f_1^3+264f_2f_1^4  
+g^2((972+516\delta_{m,0})f_2^3 \nonumber \\
&&+(504+120\delta_{m,0})f_2^2f_1+(612+300\delta_{m,0})f_2f_1^2-(216+168\delta_{m,0})f_1^3)+g^4( (504+264\delta_{m,0})f_2 \nonumber \\
&&-(144+168\delta_{m,0})f_1)\Bigr) +\frac{g}{|\alpha|} \times \begin{cases}
-(f_1+f_2) H_z^2 \ \ \ \ \   (\bm{H} \parallel \bm{z)} \\
-2f_2 H_{\theta}^2  \ \ \ \ \   (\bm{H} \parallel \bm{\theta)}
\end{cases}
=0,
\end{eqnarray}
\begin{eqnarray}
&&2\frac{\partial^2 g}{\partial \rho^2}
+\frac{1}{\rho}\left(\delta_{m,0}\frac{\partial f_1}{\partial \rho}+\delta_{m,0}\frac{\partial f_2}{\partial \rho}+2\frac{\partial g}{\partial \rho}\right)
+\frac{1}{\rho^2}\left(3\delta_{m,0}f_1-5\delta_{m,0}f_2-(8+2(m+1)^2)g\right)  
+g \nonumber \\
&&-\frac{g}{6}\left((3+\delta_{m,0})f_1^2+(3+\delta_{m,0})2f_2^2+2f_1f_2+2g^2\right)  \nonumber \\
&&-\frac{|\alpha|}{36\beta^2}\gamma \Bigl(72g^5 
+((288+120\delta_{m,0})f_1^2+(144-48\delta_{m,0})f_1f_2+(288+120\delta_{m,0})f_2^2)g^3  \nonumber \\
&&+ 
((144+72\delta_{m,0})f_1^4+(180+60\delta_{m,0})f_1^3f_2+(288+120\delta_{m,0})f_1^2f_2^2+(180+60\delta_{m,0})f_1f_2^3 \nonumber \\
&&+(144+72\delta_{m,0})f_2^4)g \Bigr) +\frac{g}{|\alpha|} \times 
\begin{cases}
0 \ \ \ \ \   (\bm{H} \parallel \bm{z)} \\
-g H_{\theta}^2/2  \ \ \ \ \   (\bm{H} \parallel \bm{\theta)}
\end{cases}
=0.
\label{eom-cyl}
\end{eqnarray}

\normalsize{
\begin{itemize}
\item  The equation of motions for $xyz$-basis ($n=0$) 
are given as follows:
\end{itemize}
}

\begin{eqnarray}
&&5\frac{\partial^2 f_1}{\partial \rho^2}-\frac{\partial^2 f_2}{\partial \rho^2}+\frac{1}{2}(-\delta_{m,2}+\delta_{m,-2})\frac{\partial^2 g}{\partial \rho^2}
+\frac{1}{\rho}\left(5\frac{\partial f_1}{\partial \rho}-\frac{\partial f_2}{\partial \rho}+\left(-\frac{3}{2}\delta_{m,2}-\frac{1}{2}\delta_{m,-2}\right)\frac{\partial g}{\partial \rho}\right) \nonumber \\
&&+\frac{1}{\rho^2}\left(-5f_1+f_2+\frac{1}{2}(-\delta_{m,2}+\delta_{m,-2})g \right)  \nonumber \\ 
&&+3f_1-f_1\left(f_1^2+f_2^2+f_1f_2+\frac{5+2\delta_{m,0}}{3} g^2\right) +f_2\left(\frac{1+\delta_{m,0}}{3}g^2\right)  \nonumber \\
&&-\frac{|\alpha|}{36\beta^2}\gamma \Bigl(216f_1^5+408f_1^4f_2+672f_1^3f_2^2+528f_1^2f_2^3+264f_1f_2^4 
+g^2((972+516\delta_{m,0})f_1^3 \nonumber \\
&&+(504+120\delta_{m,0})f_1^2f_2+(612+300\delta_{m,0})f_1f_2^2-(216+168\delta_{m,0})f_2^3)+g^4( (504+264\delta_{m,0})f_1 \nonumber \\
&&-(144+168\delta_{m,0})f_2)\Bigr)  
+\frac{g}{|\alpha|} \times \begin{cases}
-(f_1+f_2) H_z^2 \ \ \ \ \   (\bm{H} \parallel \bm{z)} \\
f_2 H_{\theta}^2  \ \ \ \ \   (\bm{H} \parallel \bm{\theta)}
\end{cases}
=0,
\end{eqnarray}
\begin{eqnarray}
&&5\frac{\partial^2 f_2}{\partial \rho^2}-\frac{\partial^2 f_1}{\partial \rho^2}+\frac{1}{2}(-\delta_{m,2}+\delta_{m,-2})\frac{\partial^2 g}{\partial \rho^2}
+\frac{1}{\rho}\left(5\frac{\partial f_2}{\partial \rho}-\frac{\partial f_1}{\partial \rho}+\left(-\frac{3}{2}\delta_{m,2}-\frac{1}{2}\delta_{m,-2}\right)\frac{\partial g}{\partial \rho}\right)  \nonumber \\
&&+\frac{1}{\rho^2}\left(-5f_2+f_1+\frac{1}{2}(-\delta_{m,2}+\delta_{m,-2})g \right)  \nonumber \\ 
&&+3f_2-f_2\left(f_1^2+f_2^2+f_1f_2+\frac{5+2\delta_{m,0}}{3}g^2 \right)+f_1\left(\frac{1+\delta_{m,0}}{3}g^2\right) \nonumber \\
&&-\frac{|\alpha|}{36\beta^2}\gamma \Bigl(216f_2^5+408f_2^4f_1+672f_2^3f_1^2+528f_2^2f_1^3+264f_2f_1^4 
+g^2((972+516\delta_{m,0})f_2^3 \nonumber \\
&&+(504+120\delta_{m,0})f_2^2f_1+(612+300\delta_{m,0})f_2f_1^2-(216+168\delta_{m,0})f_1^3)+g^4( (504+264\delta_{m,0})f_2 \nonumber \\
&&-(144+168\delta_{m,0})f_1)\Bigr)  
+\frac{g}{|\alpha|} \times \begin{cases}
-(f_1+f_2) H_z^2 \ \ \ \ \   (\bm{H} \parallel \bm{z)} \\
-2f_2 H_{\theta}^2  \ \ \ \ \   (\bm{H} \parallel \bm{\theta)}
\end{cases}
=0,
\end{eqnarray}
\begin{eqnarray}
&&2\frac{\partial^2 g}{\partial \rho^2}+\frac{1}{4}(-\delta_{m,2}+\delta_{m,-2})\left(\frac{\partial^2 f_1}{\partial \rho^2}+\frac{\partial^2 f_2}{\partial \rho^2}\right) 
+\frac{1}{\rho}\left(\left(\frac{1}{4}\delta_{m,2}+\frac{3}{4}\delta_{m,-2}
\right)\left(\frac{\partial f_1}{\partial \rho}+\frac{\partial f_2}{\partial \rho}\right) 
+2\frac{\partial g}{\partial \rho}\right) \nonumber \\
&&+\frac{1}{\rho^2}\left(
\frac{m+1}{4}(-\delta_{m,2}+\delta_{m,-2})(f_1+f_2)
-2(m+1)^2g\right)  \nonumber \\ 
&&+g-\frac{g}{6}\left((3+\delta_{m,0})f_1^2+(3+\delta_{m,0})2f_2^2+2f_1f_2+2g^2\right)  \nonumber \\
&&-\frac{|\alpha|}{36\beta^2}\gamma \Bigl(72g^5 
+((288+120\delta_{m,0})f_1^2+(144-48\delta_{m,0})f_1f_2+(288+120\delta_{m,0})f_2^2)g^3  \nonumber \\
&&+ 
((144+72\delta_{m,0})f_1^4+(180+60\delta_{m,0})f_1^3f_2+(288+120\delta_{m,0})f_1^2f_2^2+(180+60\delta_{m,0})f_1f_2^3 \nonumber \\
&&+(144+72\delta_{m,0})f_2^4)g \Bigr) +\frac{g}{|\alpha|} \times \begin{cases}
0 \ \ \ \ \   (\bm{H} \parallel \bm{z)} \\
-g H_{\theta}^2/2  \ \ \ \ \   (\bm{H} \parallel \bm{\theta)}
\end{cases}
=0.
\label{eom-xyz}
\end{eqnarray}

\end{widetext}

\bibliography{./3p2-ref.bib}

\end{document}